\documentclass[citeautoscript,floatfix,aps,prl,twocolumn,superscriptaddress]{revtex4-1}
\usepackage{graphicx}
\usepackage{amsmath}
\usepackage{amssymb}
\usepackage{dcolumn}
\usepackage[version=4]{mhchem}
\usepackage{latexsym}
\usepackage{rotating}
\usepackage{epstopdf}
\usepackage[usenames,dvipsnames]{xcolor}
\usepackage{float}
\usepackage{soul}
\usepackage{epsfig}
\usepackage{psfrag}
\usepackage{natbib}
\usepackage{bm}
\usepackage{eucal}
\usepackage{mathrsfs}
\usepackage{braket}
\usepackage{enumerate}
\usepackage{longtable}
\usepackage{bm}
\usepackage{hyperref}
\usepackage{amsfonts}
\usepackage{dcolumn}
\usepackage{bm}
\usepackage{changes}

\newcommand{\be}{\begin{equation}}
	\newcommand{\ee}{\end{equation}}
\newcommand{\bn}{\begin{eqnarray}}
	\newcommand{\en}{\end{eqnarray}}

\def\x2y2{{x^2-y^2}}

\usepackage{color} 

\usepackage{hyperref}
\hypersetup{
	colorlinks=true,final=true,
	linkcolor=red,
	citecolor=blue,
	filecolor=blue,
	urlcolor=blue,
}

\begin{document}

	\title{Vertex dominated superconductivity in intercalated FeSe}
	\author{Swagata Acharya}
	\affiliation{Institute for Molecules and Materials, Radboud University, NL-6525 AJ Nijmegen, The Netherlands}
	\affiliation{National Renewable Energy Laboratories, Golden, CO 80401, USA}
	\email{swagata.acharya@ru.nl}
	\author{Mikhail I. Katsnelson}
	\affiliation{Institute for Molecules and Materials, Radboud University, NL-6525 AJ Nijmegen, The Netherlands}
	\author{Mark van Schilfgaarde}
	\affiliation{ King's College London, Theory and Simulation of Condensed Matter,
		The Strand, WC2R 2LS London, UK}
	\affiliation{National Renewable Energy Laboratories, Golden, CO 80401, USA}
	
	\begin{abstract}

			Bulk FeSe becomes superconducting below 9\,K, but the critical temperature (T$_{c}$) is enhanced almost
        universally by a factor of $\sim$4-5 when it is intercalated with alkali elements.  How intercalation modifies
        the structure is known from in-situ X-ray and neutron scattering techniques, but why T$_{c}$ changes so
        dramatically is not known. Here we show that there is one-to-one correspondence between the enhancement in
        magnetic instabilities at certain $\bf q$ vectors and superconducting pairing vertex,
          even while the nuclear spin relaxation rate ${1}/{(T_{1}T)}$ may not reflect this
            enhancement.  Intercalation modifies electronic screening both in the plane
        and also between layers. We disentangle quantitatively how superconducting pairing vertex gains from each such
        changes in electronic screening. Intercalated FeSe provides an archetypal example of superconductivity where
        information derived from the single-particle electronic structure appears to be insufficient to account for the
        origins of superconductivity, even when they are computed including correlation effects.  We show that the
        five-fold enhancement in T$_{c}$ on intercalation is not sensitive to the exact position of the d$_{xy}$ at
        $\Gamma$ point, as long as it stays close to E$_{F}$. Finally we show that intercalation also significantly softens
          the collective charge excitations,
            suggesting the electron-phonon interaction could play some role in intercalated FeSe.

	\end{abstract}	
	\maketitle


\section*{Introduction}

With the rise of layered
materials~\cite{layeredcuo,layered2d,layeredgraphene,layeredgr,layeredibs,Kats_book,2dmater_book},
intercalation~\cite{intercalated1,intercalated2,intercalated3} and
exfoliation~\cite{exfo,exfoliation1,exfoliation2,exfoliation3,exfoliation4} became two of the most commonly used methods
to alter their structural and electronic properties.
Such intercalation with elements or molecular moieties changes the separation
between layers, screening environment and electronic structure, including Fermi surface properties.  
Indeed, shortly after their first realisation~\cite{ibs,layeredibs} in bulk crystalline form, intercalation has become one of the most popular methods to modify 
magnetic and superconducting characteristics of the iron based superconductors (IBS).

Bulk FeSe superconducts up to 9\,K, deep inside an orthorhombic phase that sets in at a much higher temperature,
90\,K~\cite{mcqueen}. The tetragonal phase, can
be made to superconduct in various ways, e.g. through doping~\cite{mizuguchi,shipra,galluzzi,sun2017,craco2014}, or
pressure~\cite{pressure,pressure1,pressure2}, as a monolayer~\cite{qing,ge2014}, or when
intercalated with alkali elements~\cite{noji,wang,potassium,barium,sodium} (Li, Na, K, Cs), under surface
doping~\cite{surface1} and ionic liquid gating~\cite{ionic1,ionic2,ionic3,ionic4}.  Remarkably,
 the critical temperature T$_{c}$ of intercalated FeSe is enhanced by roughly 4-5 times over the bulk, for many kinds
  of intercalated variants. What leads to such
dramatic enhancement in T$_{c}$ remains unanswered even after a decade of investigation.

Theoretical attempts to answer this question face some challenges as well. 
One of the primary challenges for \emph{ab-initio} theoretical calculations in such situation relates to the lack of proper information of the crystalline structure post intercalation.
In recent years, significant progress has been made on the front of determination of the crystal structure using in-situ
X-ray and neutron powder diffraction techniques~\cite{sedlmaier1,sedlmaier2,kamminga}. Further, magnetometry and
muon-spin rotation techniques are used to determine the superconducting properties of the same sample, leading to
unambiguous determination of both its structural and superconducting properties post intercalation. Reliable information
of the crystal structure helps in performing \emph{ab-initio} theoretical calculations for these materials.

FeSe is a strongly correlated material, which is reflected in its spectral properties and spin
fluctuations~\cite{coldea,yin,symmetry2021,gretarsson}. Its particularly large Hund's
coupling~\cite{medici2013,symmetry2021} drives orbital differentiation~\cite{haule2009coherence} and large electronic
vertex corrections to its two particle instabilities~\cite{symmetry2021,nematicity}. Previous theoretical
works~\cite{valenti} on intercalated FeSe connect the enhancement in T$_{c}$ to the enhancement in Fermi surface nesting
and subsequent enhancement in density of states at Fermi energy $\rho(E_{F})$. This is the usual line of argument in
superconductors where attempts have been made to connect the enhancement in T$_{c}$ to increase in $\rho(E_{F})$ in the
spirit of BCS theory~\cite{bcs}. In BCS theory T$_{c}$ has an exponential dependence on $\rho(E_{F})$. The positive
correlation of $\rho(E_{F})$ and T$_c$ was broadly discussed for both conventional superconductors such as e.g. A15 or
C15 families \cite{VIK_book} and for high-temperature cuprates \cite{HgBaCuO,vanhove}.

On the other hand, DFT calculations of intercalated FeSe finds rather weak electron
doping compared to the parent compound~\cite{valenti} and thus no significant change of $\rho(E_{F})$ is expected. We
observe the same in both DFT and many
body-perturbative approaches (quasi-particle self-consistent \emph{GW}, QS\emph{GW}~\cite{kotani}, Fig.~\ref{band}).
While QS\emph{GW} predicts $\rho(E_{F})$ to be weakly suppressed in the intercalated material, it is often noted that
FeSe is not non-magnetic, rather paramagnetic.  We take this into account by augmenting QS\emph{GW} with Dynamical Mean
Field Theory (DMFT), QS\emph{GW}+DMFT\cite{questaal-paper}.  Paramagnetic
DMFT simulates site-local magnetic fluctuations that are crucial for FeSe~\cite{dmft1,dmft2,symmetry2021,dmft3,dmft4,dmft5,nematicity}.
When QS\emph{GW} is augmented with DMFT\cite{questaal-paper}, the primary conclusion remains, namely intercalation
weakly suppresses $\rho(E_{F})$.  In absence of two-particle vertex corrections, T$_{c}$ can only drop
in such situation.  Thus, the dramatic
enhancement in T$_{c}$ can not be explained based on any theory that is reliant on electronic density of states alone,
and a primary aim of this work is to show the quintessential driving force for enhancement of T$_{c}$ originates from
the two-particle sector.  This is revealed through careful examination of the orbital, frequency and momentum
dependence of two-particle vertex functions, in magnetic and superconducting channels, and how they quantitatively
determine the nature of T$_{c}$ enhancement. 

It is not totally unexpected since even in the BCS theory the constant $\lambda$ determining T$_c$ is the product of
$\rho(E_{F})$ and the effective inter-electron interaction, and the primary reason
why common attention is focused on $\rho(E_{F})$ is that it is much easier to calculate. This simplification can be
sometimes dangerous, and as we will show here,  intercalated FeSe is a very clear example where all the
essential changes happen in the effective interaction only. In a more formal language, this effective interaction is
nothing but the vertex, that is, an essentially {\it two-particle} characteristic of correlated systems.

When materials are intercalated, the separation between active layers for superconductivity (in this case the Fe-Fe
square planes) enhances and most experimental studies attempt to establish a relation between T$_{c}$ and the separation
between layers. Such phenomenological attempts are crucial to advance the technology to the next level, where we will be
able to intercalate samples in a controlled manner to tune superconducting properties as desired. However, most such
attempts turn out to be failures. In intercalated FeSe, there are several samples with very large inter-layer separation
where T$_{c}$ nevertheless remains invariant.  That being said, when
the inter-layer separation enhances it reduces significantly the electronic screening and that leads to larger Hund's
coupling.  This aspect of the correlated Fe-3d Hamiltonian is often entirely discarded from theoretical analysis, but it
turns out to be crucial.  Within our \emph{ab-initio} constrained QS\emph{GW}-RPA theory, we compute the changes to the
bare two-particle parameters entering into the Anderson impurity model, namely the Hubbard \emph{U} and Hund's \emph{J},
in an unbiased fashion, and show how they affect T$_{c}$.

\begin{figure*}[ht!]
	
	  \includegraphics[width=0.99\textwidth]{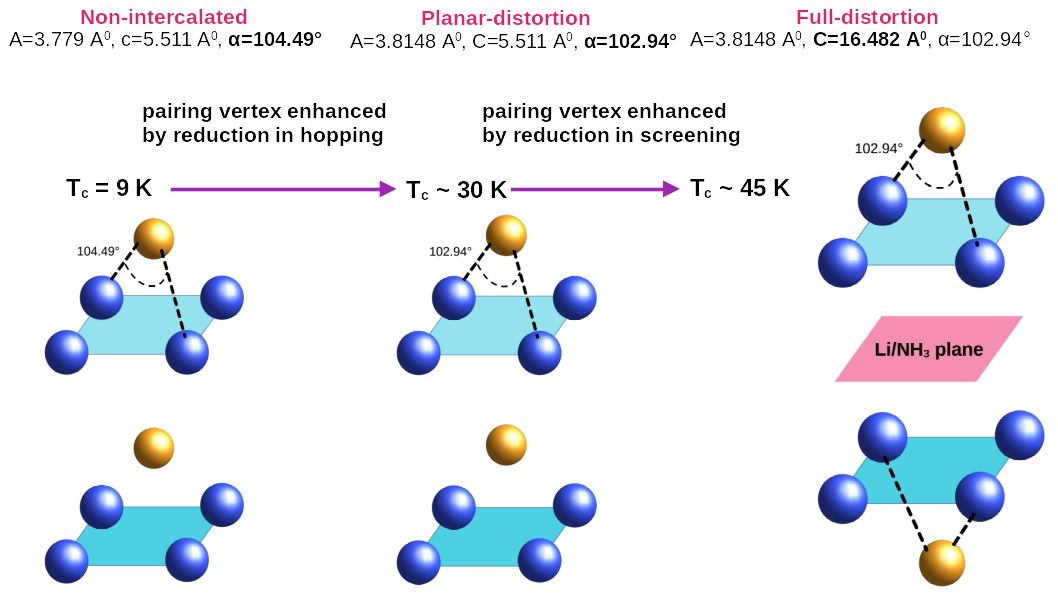}
	\caption{{\bf Separability of conditions for dramatic enhancement in T$_{c}$ on intercalation:} A two-step increment in
		T$_{c}$ on intercalation is explored; the first is due to smaller Fe-Se-Fe bond angle that reduces the Fe-Se-Fe
		hopping and makes Fe 3d states more correlated, and the second is due to enhanced \emph{c}-axis length that reduces
		electronic screening and enhances correlations. The rest of the letter discusses how these two mechanisms strongly
		modify the pairing vertex while the one-particle density of states remain nearly unchanged.}
	\label{summary}
\end{figure*}

We quantitatively analyse the consequences of the following: changes to the Fe-Se-Fe bond angles, changes to the Hund's
\emph{J} driven by enhancement in layer separations in building the correlated Hamiltonian for intercalated FeSe (see
Fig.~\ref{summary}).  We solve the correlated two-particle Hamiltonian with Bethe-Salpeter equations in the magnetic and
superconducting channels to show in detail how the five-fold T$_{c}$ enhancement in Li/NH$_{3}$ intercalated FeSe
originates.  We analyse the parent non-intercalated bulk tetragonal FeSe ($n{-}i$), a bulk FeSe structure simulated only
using the planar distortion ($p{-}d$) parameters (Fe-Se bond lengths and Fe-Se-Fe bond angles reported in the
supplemental materials Table 2 of Ref~\cite{sedlmaier2}); and also intercalated FeSe with the full structural ($f{-}d$)
distortion as reported in the same work.  The primary difference between the
planar-distortion ($p{-}d$) and full-distortion ($f{-}d$) is the enlargement of the \emph{c} axis, that reduces the
electronic screening perpendicular to the Fe-Fe square plane (see Fig.~\ref{summary}), leading to about 10\% enhancement
to the Hubbard parameters for the Fe-3d orbitals (in $n{-}i$ and $p{-}d$, \emph{U}=3.5 eV and \emph{J}=0.6 eV and in
$f{-}d$, \emph{U}=3.76 eV and \emph{J}=0.68 eV). While several prior studies have
   explored the relation between T$_{c}$ and structural parameters in iron-based
  superconductors~\cite{ref16,ref17,ref18,ref19}, here we are able to disentangle the effects coming from planar
  distortions and inter-layer separations, and show quantitatively how each of these structural changes modifies the
  strength of the pairing instability.  Changes in crystal structure leads to changes in electronic screening which
affects both the single-particle and two-particle sectors, and the enhancement in T$_{c}$ can only be explained when
both are treated in the presence of higher-order vertex corrections. Previous works have analysed the
  screened Coulomb vertex corrected pairing instabilities. in a one-band model for cuprates~\cite{ref20} and a two-band
  model for nickelates~\cite{ref21}. However, our solutions for Bethe-Salpeter equations in pairing channel keeps full
  orbital-, momentum- and two-frequency-dependence (fermionic Matsubara frequencies) of all five Fe-3d correlated
  orbitals. The only approximation we make is to put the center-of-mass co-ordinates to zero (center-mass momentum $q$=0
  and the bosonic frequency $\Omega$=0) since the linearised
    Eliashberg equation simplifies to an eigenvalue problem that can be solved then at different temperatures in the
  normal phase looking for all possible pairing instabilities and their competitions~\cite{swag19,prl2020} in a fully
  unbiased fashion.

\section*{Results and Discussion}
\begin{figure*}[ht!]
	
		  \includegraphics[width=0.99\textwidth]{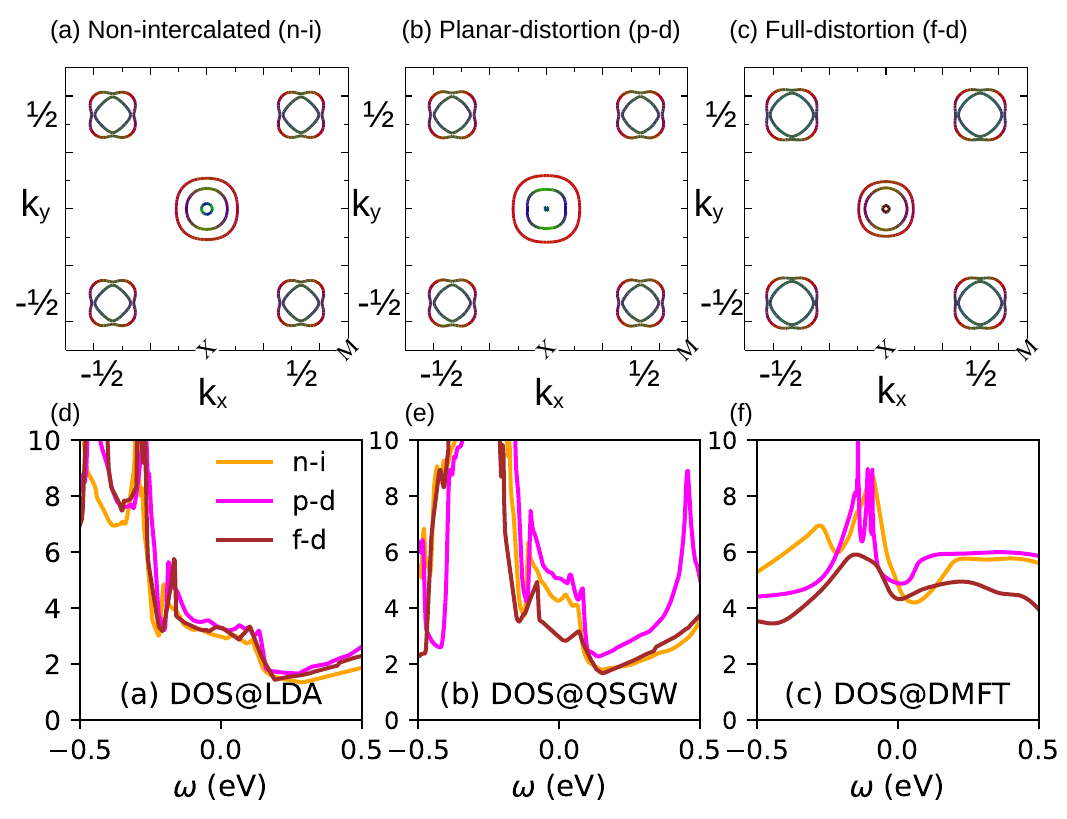}
	\caption{{\bf Failure of one-particle properties from LDA, QS\emph{GW} and QS\emph{GW}+DMFT in explaining the
			enhancement in T$_{c}$}: The orbital projected electronic QS\emph{GW} Fermi surfaces for (a) $n-i$ FeSe, (b) $p-d$
		FeSe and (c) $f-d$ FeSe are shown. The Fe-3d$_{x^2-y^2}$+d$_{z^2}$ orbitals are shown in blue, d$_{xz,yz}$ in green and
		d$_{xy}$ in red. Finally, the total density of states (d-f) from different levels of the theory are shown. In
		QS\emph{GW}+DMFT the total density of states decrease in intercalated sample, and yet T$_{c}$ enhances. k$_{x}$,k$_{y}$ are in units of $\frac{2\pi}{a}$.}
	\label{band}
\end{figure*}

We start by discussing the one-particle properties: the electronic band structure, density of states (DOS) and the Fermi
surfaces computed using LDA and QS\emph{GW} (see Fig.~\ref{band}). $\rho({E_{F}})$ enhances slightly going from $n{-}i$
to $p{-}d$, since the Fe-Se-Fe bond angle $\alpha$ reduces slightly, leading to reduced
Fe-Se-Fe hopping and narrower Fe bands.  However, the effect is nullified in $f{-}d$ owing to the enlarged
 \emph{c}-axis (see Fig.~\ref{band}(d-f)).  In this case $\rho(E_{F})$ decreases.  The
electron pockets at M becomes slightly larger, while the hole pockets at $\Gamma$ become
smaller, leading to net electron doping of the system (see Fig.~\ref{band} (a-c)).  Nevertheless, we find that the electron doping within the QS\emph{GW} approximation
remains as small as $\sim$1\%. The situation is
similar in QS\emph{GW}+DMFT where $\rho(E_{F})$ in $f{-}d$ also drops relative to $n{-}i$ and $p{-}d$.  Thus, the effective electron doping of the
system remains rather weak.  If the vertex were to remain constant,  a
five-fold enhancement in T$_{c}$ would require about 80\% enhancement in electronic density of states in a BCS picture
(T$_{c,\mathrm{BCS}}\sim\exp{[-1/V\rho(E_{F})]}$), (assuming the correlation parameters $V$ remain unchanged).

\begin{figure*}[ht]
	
		  \includegraphics[width=0.99\textwidth]{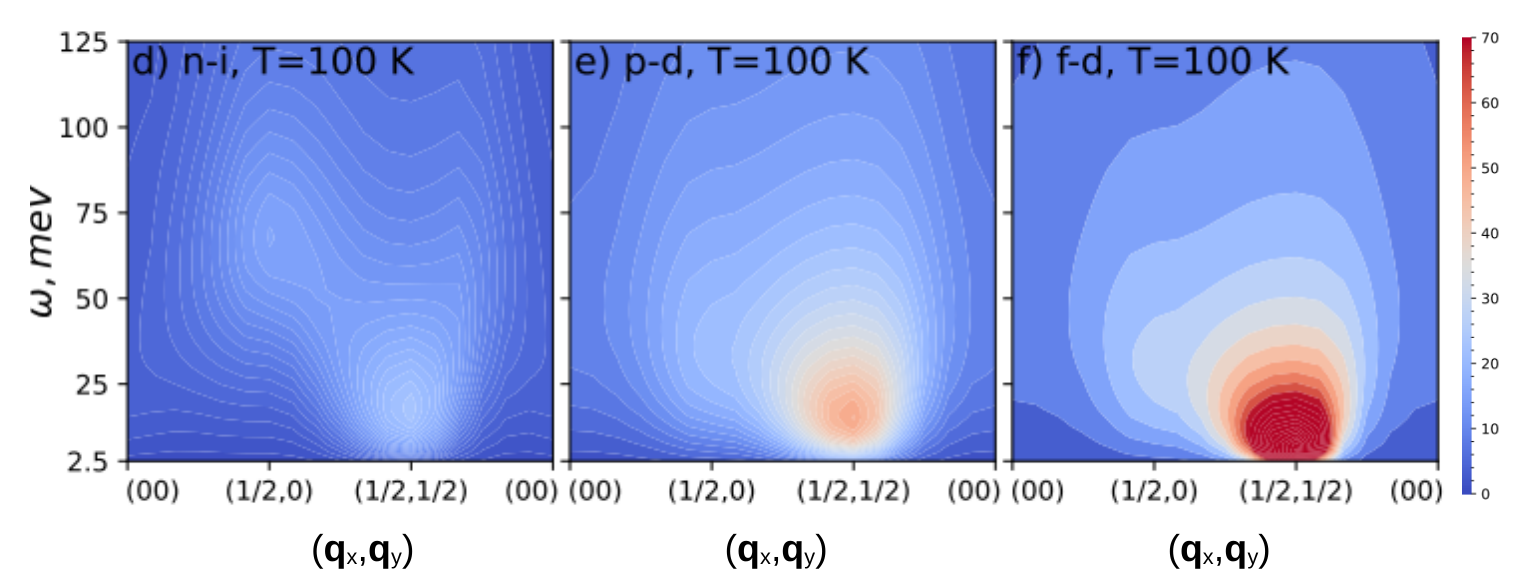}
	\caption{{\bf Enhancement in low energy magnetic glue for
			superconductivity on intercalation:} The vertex corrected dynamic and momentum resolved magnetic
		susceptibility Im$\chi^{m}(\omega,\bf q)$ is shown for (a) $n{-}i$ , (b) $p{-}d$ and (c) $f{-}d$ at 100\,K. On
		intercalation, Im$\,\chi^{m}(\omega,q)$ becomes more intense at low energies, particularly at $\bf q=(1/2,1/2)$
		which corresponds to the anti-ferromagnetic instability vector in 2-Fe atom unit cell of FeSe. Also with
		lowering of temperature the low energy structure of Im$\chi^{m}(\omega,\bf q)$ becomes more prominent.}
	\label{chi}
\end{figure*}

We now turn our attention to two-particle instabilities, particularly the frequency and momentum resolved magnetic
susceptibility Im$\,\chi^{m}(\omega,\bf q)$ (Fig.~\ref{chi}).  It is computed by solving a non-local Bethe-Salpeter
equation (BSE) that dresses the non-local polarisation bubble in the particle-hole magnetic channel by the local
irreducible dynamic vertex~\cite{nematicity,swag19}.  Both the inputs for the BSE, the single-particle vertex which
enters the self-energy and the two-particle vertex entering into response functions, are computed using DMFT.  Our
computed Im$\,\chi^{m}(\omega,\bf q)$ has been rigorously benchmarked against Inelastic Neutron Scattering
measurements~\cite{wang} in a prior study~\cite{nematicity} over all relevant energies and momenta. We
  analyse Im$\,\chi^{m}(\omega,\bf q)$ along the reciprocal lattice vectors $\bf q$=($\bf
  q_{x},q_{y},q_{z}$), in units of ${2\pi}/{a}$. We fix $q_{z}$ at 0 and explore the susceptibilities in the
  $\bf q$=($\bf q_{x},q_{y}$) plane. In the vicinity of $\bf q{=}$(1/2,1/2), intercalation ($p{-}d$ and $f{-}d$)
significantly enhances the strength of Im$\chi^{m}(\omega,\bf q)$ at lower energies, making spin-fluctuation mediated
superconductivity more favourable.  We also observe that this tendency is independent of temperature; nevertheless, with
lower temperatures the strength of spin-glue at low-energies for superconductivity becomes more prominent.  To
understand the essential ingredient of such spin susceptibilities in the static limit, we recompute
$\chi^{m}(q)=\chi^{m}(\omega{=}0,\bf q)$ with a vertex and without (RPA).

\begin{figure*}[ht]
		  \includegraphics[width=0.99\textwidth]{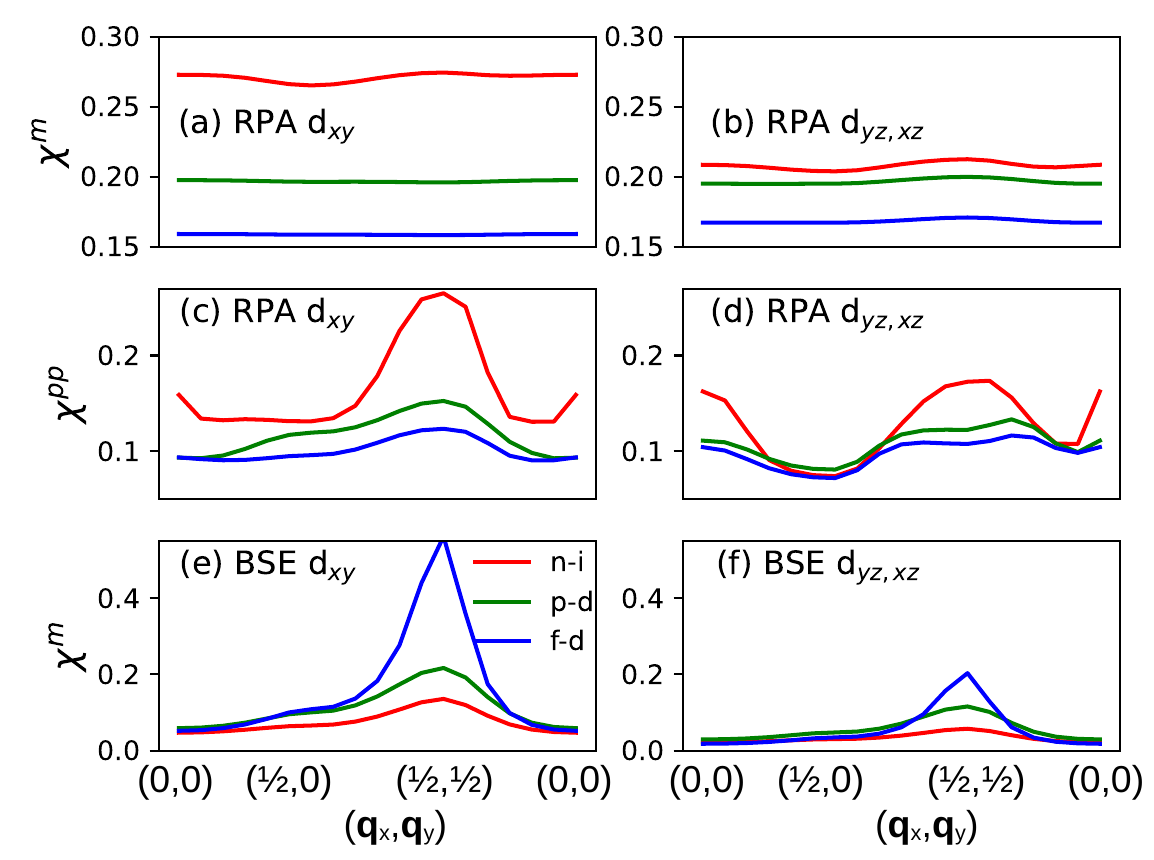}
	\caption{{\bf Shortcoming of the RPA theory for magnetic and superconducting susceptibilities:} $\chi^m$ and
		$\chi^{pp}$ are resolved in different intra-orbital channels. RPA theory predicts magnetic (a,b) and superconducting (c,d)
		susceptibilities to be maximal in the non-intercalated sample, while vertex corrections predict a nearly
		five-fold enhancement in magnetic instability (e,f) in the intercalated sample.}
	\label{suscep}
\end{figure*}

Within the RPA, $\chi^{m}(\bf q)$ is resolved in different intra-orbital channels (Fig.~\ref{suscep}(a,b)) and
it remains the largest in $n{-}i$ at all $\bf q$.  Intercalation weakly suppresses $\chi^{m}(\bf q)$ in different inter-orbital
channels suggesting a weak suppression in superconductivity in a purely RPA picture of spin-fluctuation mediated
superconductivity. Our computed $\chi^{m}(\bf q)$ has similarities with the susceptibility computed in Ref.~\cite{valenti}.
Particularly striking is the very large intra-orbital elements of $\chi^{m}(\bf q)$ at $\bf q{=}0$.  This is primarily a
shortcoming of the RPA approximation: without vertex corrections the momentum dependence of $\chi^{m}(\bf q)$
is rather weak and there is no clear separation between $\chi^{m}$ at $\bf q{=}$0 and $\bf q{=}$(1/2,1/2).  However,
experimentally, in bulk FeSe magnetic fluctuations at $\bf q{=}$0 are rather weak compared to the main instability at
$\bf q{=}$(1/2,1/2)~\cite{wang} and it is the vertex that brings in the
needed momentum dependent variation in $\chi^m$, as we have shown previously~\cite{nematicity}.

We compute the RPA particle-particle superconducting susceptibility $\chi^{pp}(\bf q)$, and resolve it by intra-orbital
components (Fig.~\ref{suscep}(c,d)).  Indeed $\chi^{pp}(\bf q)$ gets weakly reduced on intercalation. This conclusion
appears qualitatively slightly different from earlier work~\cite{valenti}, and the main reason for that, we think, is in
how the RPA pairing susceptibility was computed.  In Ref.~\cite{valenti} it was computed from a tight-binding model
derived from DFT, while in the present case the eigenfunctions are computed directly from QS\emph{GW}+DMFT.  From this
earlier work~\cite{valenti} it can also be seen that the superconducting eigenvalue $\lambda$ gets weakly reduced for
most of the doping range, and gets weakly enhanced at the maximal electron doping.  However, the true intercalated
sample in experiment does not correspond to the maximal doping and it is rather difficult to understand from
their figure whether $\lambda$ for the experimentally intercalated sample should increase or decrease.  In any case,
such minor differences between these two different RPA calculations are not relevant for the essential fact of a
 five-fold enhancement in T$_{c}$ on intercalation.

\begin{figure*}[ht]
	
		  \includegraphics[width=0.99\textwidth]{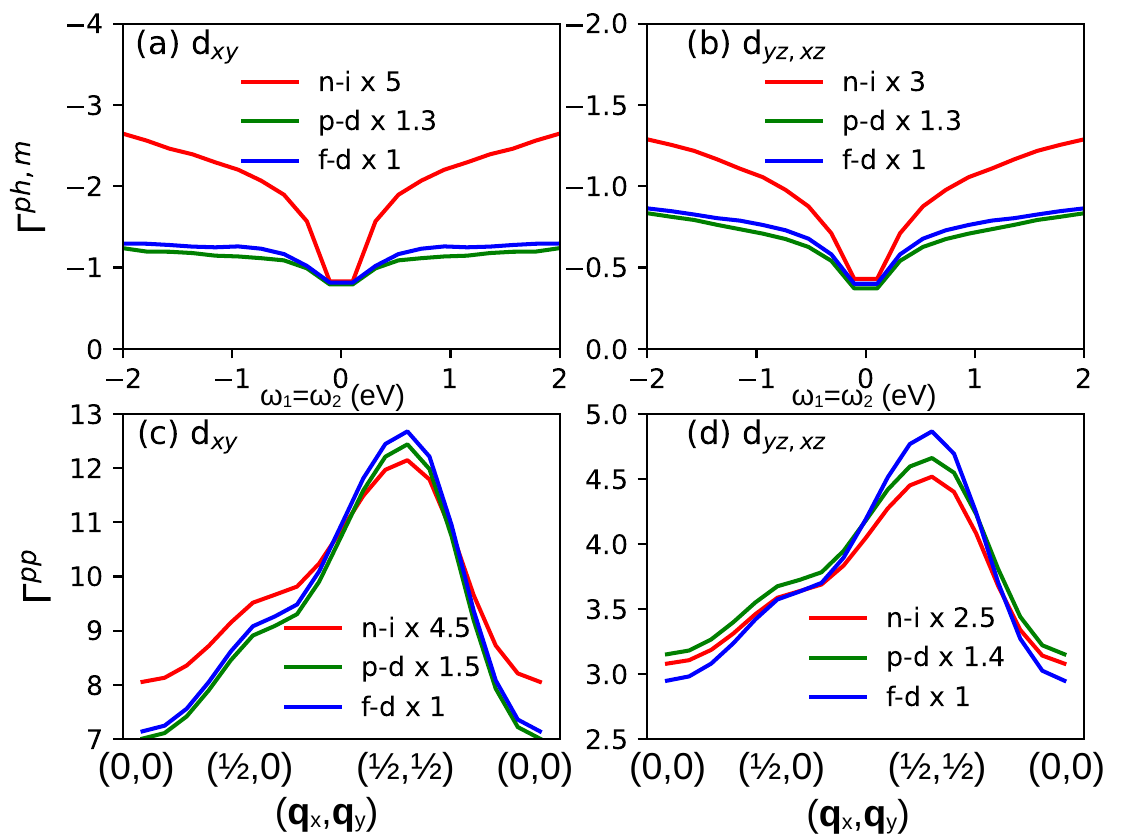}
	\caption{{\bf Vertex mediated five-fold enhancement in T$_{c}$ on intercalation:} The orbital projected magnetic
		$\Gamma^{ph,m}$ (a,b) and pairing $\Gamma^{pp}$ (c,d) vertex functions show a factor of $\sim$5 enhancement in the d$_{xy}$
		channel, while a factor of $\sim$3 enhancement in the d$_{yz,xz}$ channels on intercalation. The enhancement in
		T$_{c}$ directly correlates with enhancement in pairing vertex strength in the d$_{xy}$ channel.}
	\label{vertex}
\end{figure*}

When $\chi^{m}(\omega,\bf q)$ is calculated with the vertex, it shows a systematic enhancement on intercalation
(Fig.~\ref{suscep}(e,f)). The static $\chi^{m}(1/2,1/2)$ in the d$_{xy}$ channel gets enhanced by a factor of nearly 5,
and $\sim$3 in the d$_{yz,xz}$ channels. However, $\chi^{m}_{xy}(1/2,1/2)$ always remains at least a factor of two
larger than $\chi^{m}_{yz,xz}(1/2,1/2)$, suggesting that the low energy spin
fluctuations originate primarily in the d$_{xy}$ channel.  Orbital-differentiation is a signature of Hund's
correlations; mass enhancement factors can be quite different for different
orbitals~\cite{dmft1,dmft2,dmft3,nematicity,symmetry2021,kostin2018}.  The effect on $\chi$ is a two-particle analogue
of the orbital-differential for the self-energy.  The fact that d$_{xy}$ is the
primary source of magnetic fluctuations in a range of strongly correlated chalcogenide and pnictide superconductors was
discussed in previous works~\cite{symmetry2021,nematicity,prl2020}.  To understand the role of the vertex functions in
this remarkable enhancement in $\chi^{m}(\bf q)$ we analyse the magnetic vertex functions $\Gamma^{ph,m}$
(Fig.~\ref{vertex}(a,b)) and their energy, momentum and orbital dependence. $\Gamma^{ph,m}
(\omega_{1},\omega_{2},\Omega$) is a dynamic quantity that depends on two Matsubara frequency indices
($\omega_{1,2}$) and one bosonic frequency ($\Omega$). We observe that the five-fold enhancement in $\chi^{m}$ is
directly related to the five-fold enhancement in $\Gamma^{ph,m}$ in the d$_{xy}$ channel in the static limit
($\Omega{=}0$ and $\omega_{1}{=}\omega_{2}$).  $\Gamma^{ph,m}$ also gets enhanced on intercalation in the d$_{yz,xz}$
channels but only by a moderate amount.  Also, the magnetic vertex corrections always remain about factor of two larger
in the d$_{xy}$ channel than the d$_{xz,yz}$ channels, consistent with the magnetic susceptibilities.  This shows
that the enhancement in magnetic fluctuations at low energies on intercalation is purely a phenomenon emerging from the
two-particle electronic vertex, and it is not contained in the bare RPA polarizability, even when computed
using the Green's functions G$_{k,\omega}$ dressed with the DMFT self-energy $\Sigma(\omega)$.

We further analyse the momentum and orbital structure of the pairing vertex $\Gamma^{pp}$ at $\Omega{=}0$ (after all
internal frequencies are integrated). In complete consistency with $\chi^{m}(1/2,1/2)$ we observe a similar nearly
five-fold enhancement in $\Gamma^{pp}(q{=}1/2,1/2)_{xy}$ (see Fig.~\ref{vertex}(c,d)). Weaker enhancements in the
d$_{yz,xz}$ pairing vertex can be observed as well. Intriguingly enough, $\Gamma^{pp}(q{=}1/2,1/2)_{xy}$ gets enhanced
in $p{-}d$ compared to $n{-}i$ by a factor of $\sim$3, and this is purely due to the changes in Fe-Se-Fe bond angle. The
reduction in screening from the increased
\emph{c}-axis length that enhances \emph{U},\emph{J} by about 10\% accounts for about a third of the total
enhancement in T$_{c}$ from $\sim$9 K to $\sim$45 K. We also observe that a proper treatment of the $\Gamma^{pp}$
reduces the contribution to superconducting instability originating from $\bf q{=}$0 and makes it mostly dominated by
$\bf q{=}$(1/2,1/2). Further, we compute the superconducting order parameters and find that while the leading instability is
of extended \emph{s}-wave in nature~\cite{nematicity,symmetry2021} in $n{-}i$ and $f{-}d$, it has d$_{x^2-y^2}$ symmetry
in $p{-}d$. This is another testimony to the importance of reliable computation of vertex and Hubbard correlation
parameters in such strongly correlated systems, where moderate changes to these quantities can lead to significant
qualitative changes to collective instabilities.

\subsection*{Missing d$_{xy}$ hole pocket and the superconducting instability}

We next address the major limitation of the present \emph{ab initio} theory, namely its prediction of a
  hole pocket in the d$_{xy}$ channel at the $\Gamma$ point, which is missing in photoemission
  experiments~\cite{watsonprb,watsonnjp}, this band is found to fall slightly below $E_{F}$.  As we have discussed
  elsewhere~\cite{symmetry2021}, the Fermi surfaces computed from our QS\emph{GW}+DMFT approach, significantly reduces
  the hole pocket compared to DFT+DMFT approaches but it does not drive the d$_{xy}$ state below $E_{F}$.  The origin
  of this discrepancy originates from a non-local self-energy, likely magnetic fluctuations or the electron-phonon
  interaction.  Neither of these are yet built into the present theory, but whatever the cause it is important to
  assess its effect on the conclusions of this paper.  The absolute position of the d$_{xy}$ states at $\Gamma$
  point can be sensitive to doping~\cite{ref11,ref12}, intercalation and other structural changes.
	
\begin{figure*}[ht]
	
		  \includegraphics[width=0.99\textwidth]{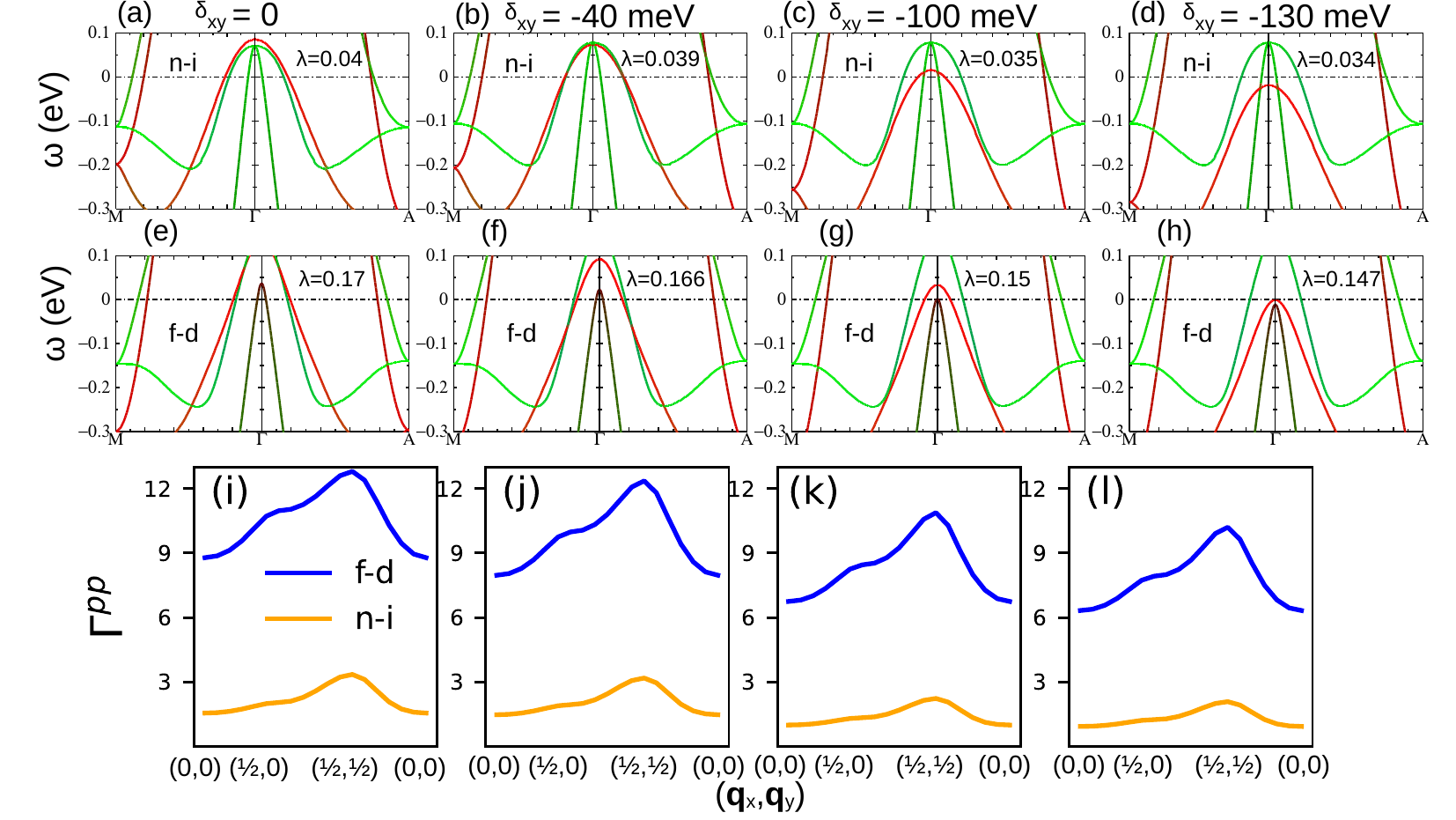}
	\caption{{\bf d$_{xy}$ hole pocket and its impact on superconductivity:} For both non-intercalated (a-d) and intercalated (e-h) the electronic band structure is shown along M-$\Gamma$-A path for different values of shift $\delta_{xy}$ in potential. Red represents the d$_{xy}$ orbital character and green represents d$_{xz,yz}$ orbital characters. The leading superconducting eigenvalues $\lambda$ remains roughly 4-5 times larger in intercalated samples for all  $\delta_{xy}$. The pairing vertex $\Gamma^{pp}$ uniformly enhances by similar factors in intercalated variants for all such $\delta_{xy}$, as the electron pockets around M and A, always contain significant d$_{xy}$ component.}
	\label{potential}
\end{figure*}

Here we model how the role of proximity of d$_{xy}$ state to E$_{F}$ affects superconductivity by adding
  an external potential to shift to the d$_{xy}$ state only, leaving the remainder of the system intact.  It is similar
  in spirit to QS\emph{GW}+U+DMFT where U is applied only to the d$_{xy}$ orbital. We use U as a free parameter to
  create a potential shifts $\delta_{xy}$ to d$_{xy}$ and explore its consequence in superconducting pairing instability
  by solving the BSE in the pairing channel, exactly as before. Here we show results for four cases, with $\delta_{xy}$
  of 0, $-$40, $-$100 and $-$130 meV (see Fig.~\ref{potential} (a-d)). In the first three cases, (a)-(c), the d$_{xy}$ hole
  pocket survives, although it keeps getting smaller with larger $\delta_{xy}$. We find that the leading eigenvalue
  $\lambda$ for superconductivity in $s^\pm$ channel remains nearly invariant for $\delta_{xy}$=$-$40 and $-$100 meV where
  the d$_{xy}$ hole pocket still survives.  When finally $\delta_{xy}$=$-$130 meV, the d$_{xy}$ hole pocket is pushed
  below $E_{F}$ by $\sim$25\,meV, close to the position observed in photoemission.  In that case we find $\lambda$
  is reduced only slightly compared to $\delta_{xy}$=0. However, note that in all cases in Fig.~\ref{potential}
  (a-d), the d$_{xy}$ character survives at the electron pockets. Superconductivity is a low energy phenomena and it is
  known that for multi-orbital superconductivity, it is important that both the narrow and dispersive orbitals are
  present close to the Fermi energy and it is not necessary that all have them have to be right at
  $E_{F}$~\cite{ref13,ref14,ref15}. In some of our older works, we discussed the impact of anomalous screening on properties like, Kondo physics, Mott transitions and superconductivity, even when the van Hove singularities in electronic density of states sit beyond the thermal broadening energy scales from the Fermi surface~\cite{mik1,mik2}. Our fully \emph{ab-initio} framework establishes that the proximity of the most
  correlated d$_{xy}$ state to the Fermi energy, is key to superconductivity~\cite{symmetry2021}. If it gets pushed too
  far below the Fermi energy, the superconducting instability is suppressed. This is understandable since the paramagnon
  dispersion in bulk FeSe is $\sim$100 meV and the superconducting gap energy scale is only about 1 meV. Bands that are
  pushed far beyond these typical relevant energy scales, would have less impact on the pairing mechanism.

We now explore the consequence of similar shifts $\delta_{xy}$ in intercalated (f-d) FeSe. Note that in
  Fig.~\ref{suscep} and Fig.~\ref{vertex}, we have shown that for $\delta_{xy}$=0, intercalation (f-d) produces a
  $\sim$4-5 factor enhancement is pairing instability compared to the non-intercalated (n-i) bulk FeSe. As we add,
  $\delta_{xy}$=$-$40, $-$100 and $-$130 meV (see Fig.~\ref{potential}(f-h)), we find that in all cases the relative pairing
  strengths still enhances $\sim$4-5 on intercalation. This intriguing result supports the essential claim of our work
  that the nearly five-fold T$_{c}$ enhancement on intercalation is a robust fact and is not sensitive to the presence
  or absence of the d$_{xy}$ hole pocket. This is possible because d$_{xy}$ orbital character is still present in the
  electron pockets and the pairing instability mediated by d$_{xy}$ still gets enhanced by a similar factor on
  intercalation.  A significant challenge for future \emph{ab-initio} studies will be to explain the physical mechanism
  that slightly suppresses the d$_{xy}$ hole pocket.

\subsection*{Different experimental observations in intercalated FeSe}

Experiments offer valuable insights into the origins of superconductivity, and ideally experiments
  such as ARPES, NMR and Knight shift could provide hard tests that either lend support to the conclusions we have
  drawn, or be at variance with them.  We present here some key experimental findings that broadly support the
  theoretical findings; however there are enough gaps in both experiment and theory that some caution is needed: we
  cannot exclude the possibility that a boson we have not considered, (e.g. the electron-phonon interaction) also makes
  some contribution to superconductivity in intercalated FeSe.  For example, an ARPES study on Li-intercalated
  FeSe~\cite{arpes} at inside the superconducting phase at 20K suggests that all the hole pockets are pushed slightly
  below E$_{F}$.  This study, as well as valence analysis of Fe~\cite{sedlmaier1,sedlmaier2}, indicate that the
  intercalated samples are slightly electron-doped.  Even in intrinsic FeSe d$_{xy}$ state is below
  E$_{F}$~\cite{watson2016,watson2017} and d$_{xz,yz}$ only leads to a very small hole pocket.  Some experiments suggest
  suppression of the hole pocket suppresses superconductivity, in accord with a traditional nesting picture.  Two
  instances of heavily electron-doped systems that do not superconduct are Fe$_{1.03}$Se~\cite{imai2009} and
  Fe$_{0.9}$Co$_{0.1}$Se.  Li-intercalated FeSe~\cite{arpes}, on the other hand are the counterexamples (alkali doped FeSe~\cite{alkaliarpes1,alkaliarpes2,alkaliarpes3} and intercalated FeSe) that establishes a
  hole pocket crossing $E_{F}$ at $\Gamma$, and therefore a static one-particle nesting picture, is not essential.  Our
  theoretical treatment indicate that provided those hole pockets lie close enough to E$_{F}$ they can mediate strong
  pairing even if they do not cross it.  A similar argument was made based on a two-band model
  Hamiltonian~\cite{incipient}, where the authors showed that spin fluctuations can mediate pairing in the strong
  coupling limit when the electron-like pocket remains at E$_{F}$ while the hole-like pocket becomes `incipient'.  This
  suggests that the traditional static concept of nesting needs to be generalised to a dynamical one, where the
  frequency-dependence of $\chi$ plays a key role.

\begin{figure*}[ht]
	
		  \includegraphics[width=0.99\textwidth]{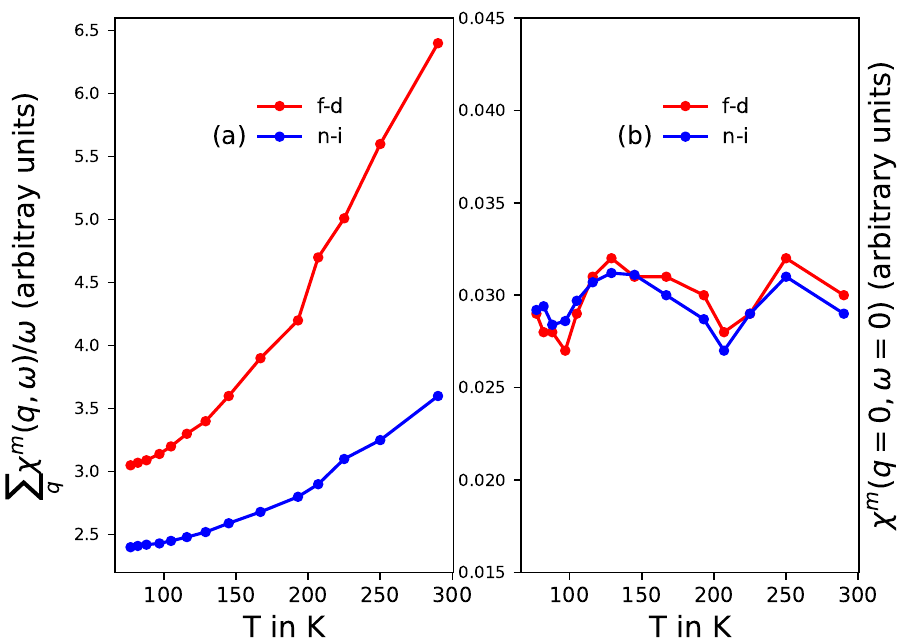}
	\caption{{$\bf q$-integrated and $\bf q$=0 spin susceptibilities for comparison against the NMR $\frac{1}{T_{1}T}$ and Knight Shift data:} DMFT vertex corrected $\sum\frac{\mathrm{Im}\chi^m(\bf q,\omega)}{\omega}$ and $\chi(\mathbf{q}{=}0, \omega{=}0)$  are computed in the temperature range 300 K to 77 K. (a) While $\sum\frac{\mathrm{Im}\chi^m(\bf q,\omega)}{\omega}$ remains almost twice large in the intercalated phase at high temperatures compared to the non-intercalated phase, at lower temperatures it is only about 20\% larger. (b) $\chi^m(\mathbf{q}{=}0, \omega{=}0)$ remains almost invariant over all temperatures between the non-intercalated and intercalated phases.}
	\label{nmr}
\end{figure*}

Other key experiments are NMR and the Knight shift.  We compute $\sum{\mathrm{Im}\,\chi^m(\bf
    q,\omega)/{\omega}}$, which is the main factor in determining ${1}/{(T_{1}T)}$ measured by NMR; and also
  $\chi^m(\mathbf{q}{=}0,\omega{=}0)$, which is the main factor controlling the Knight shift K$_{S}$.  Fig.~\ref{nmr}(a)
  shows that $\sum{\mathrm{Im}\,\chi^m(\bf q,\omega)/{\omega}}$ for f-d and n-i are differ widely at room temperature,
  but the difference shrinks with temperature, to about 20\% at 77 K.  This is consistent with our interpretation that
  while at a particular $\bf q$ vector $\chi^m$ can get enhanced by a factor of 4-5 in the f-d, the local quantity may
  not show similar enhancement.  Turning to the Knight shift, the theory predicts (see Fig.~\ref{nmr}(b)) almost no difference between f-d and
  n-i in $\chi^m(\mathbf{q}{=}0,\omega{=}0)$ at any temperature.  Both of these observations are found in experiments as
  well~\cite{imai2009,nmr}.  Together, they suggest that our computed magnetic susceptibilities and their momentum and
  energy structures are of good fidelity and reasonably consistent with NMR data.  However, we can not fully interpret the absence of the build up of ${1}/{(T_{1}T)}$ right above T$_{c}$ as observed in experiments. We believe, to an extent we understand why that is the case, but a complete understanding is lacking. We discuss this in detail in the next paragraph. As an additional caveat, the QMC solver available
  to us limits the temperatures we can reach.  The lowest temperature for which we could compute vertex corrected
  susceptibilities was 77 K, somewhat above the critical region around 45 K.

In bulk FeSe, the primary nesting vector is (1/2,1/2) (in the unit cell that contains two Fe atoms).  The primary nesting
  vectors are different in FeSe and FeSe$_{1-x}$Te$_{x}$~\cite{fesete}. With Te doping more than one nesting vector
  emerges. A similar situation occurs in alkali doped FeSe~\cite{alkali} where the primary nesting vectors are entirely
  different~\cite{alkaliq,alkaliq1} from bulk undoped FeSe. A new primary nesting vector, often, does co-exist with the old one and also
  with other additional nesting vectors (several vectors with similar strengths of spin fluctuations). What it primarily suggests is that spectral weights get redistributed in the material over different $\bf q$ and energies, on doping. One can imagine that what the material chooses as the primary nesting vector from many, in doped
  systems, can depend on multitude of factors and the balance can shift easily (depending on parameters like atomic co-ordinates and changes in hopping parameters). Intriguingly enough, in almost all alkali
  doped FeSe compounds, the hole pockets appear to be pushed below
  E$_{F}$~\cite{alkaliarpes1,alkaliarpes2,alkaliarpes3}, much like its intercalated counterpart. This is of extreme
  relevance to our case, because ${1}/{(T_{1}T)}$ is a local quantity that sums over dynamical spin susceptibility
  over all $\bf q$ (in the $\omega\rightarrow{0}$ limit). In complicated cases like these, where there are spectral weight redistribution over various $\bf q$
  points, the summed over quantity over a certain temperature window may still appear very similar from different
  materials (for example, in bulk FeSe, alkali doped FeSe~\cite{alkalinmr} and intercalated FeSe~\cite{nmr}). Also, when
  we check the magnetic susceptibility $\chi^m$ at the vector $\bf q$=(1/2,1/4,1/2) (unfolded) we do find a strong enhancement
  in spin fluctuations compared to non-intercalated case. This is the same vector where the primary nesting appears when
  FeSe is doped with Rb or K. Additionally, from our ${1}/{(T_{1}T)}$ calculations we also observe that as we include more q-points to compute  $\sum{\mathrm{Im}\,\chi^m(\bf
    q,\omega)/{\omega}}$, the curves for the n-i and f-d phases come closer at lower temperatures. This is a clear indication for the fact that when spin fluctuations are distributed over several $\bf q$ vectors we need to be more careful about obtaining this quantity. A counter-argument to this could be, APRES suggests that in (alkali doped and) intercalated samples nesting from (1/2,1/2) is removed and data from Neutron scattering suggests that a new nesting vector appears at (1/2,1/4,1/2), then would not it be reasonable to argue that magnetic fluctuations and superconductivity chooses this new vector (1/2,1/4,1/2), instead of (1/2,1/2)? The problem with this argument is that if this new vector nests the Fermi surface and mediate spin fluctuations it still does not answer why the build up in ${1}/{(T_{1}T)}$ remains absent right above T$_{c}$. This gives us more confidence in our above analysis, that we need to be more careful in interpreting ${1}/{(T_{1}T)}$, in cases where multiple (finely balanced) nesting vectors can emerge under doping or intercalation. When we add to this our observation that superconductivity does not necessarily gain from a static nesting picture, but rather from `incipient' bands, we would have to be even more careful in both computing and interpreting  ${1}/{(T_{1}T)}$, since the relevant `incipient' bands for superconductivity are sitting few meV below E$_{F}$. (Note that a similar situation occurs in uni-axially strained Sr$_{2}$RuO$_{4}$, where a new spin
  susceptibility peak emerges at $\bf q$=(0.5,0.25,0)~\cite{sr2ruo4} while the peak at incommensurate $\bf
  q$=(0.3,0.3,0) from the unstrained material also survives with almost equal intensity and the T$_{c}$ enhances on
  strain.)

\subsection*{Softening of collective charge excitations and possibility of divergence in electron-phonon vertex}

\begin{figure*}[ht]
	
		  \includegraphics[width=0.99\textwidth]{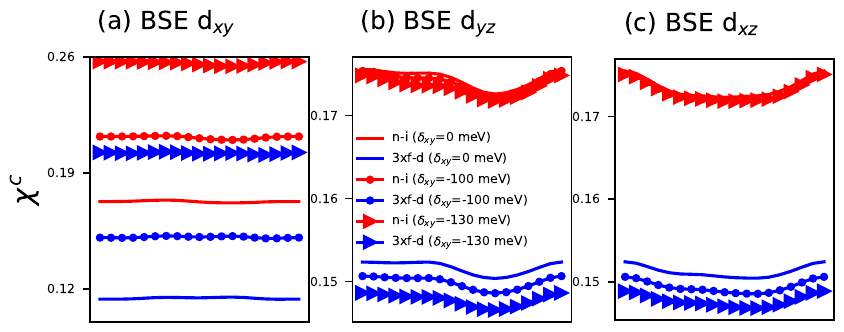}
	\caption{{\bf Softening of charge susceptibility $\chi^c$ on intercalation:} DMFT vertex corrected $\chi^c$ is plotted in different intra-orbital channels. For all the channels real part of $\chi^c$ reduces on intercalation. We plot three times the $\chi^c$ for the intercalated phase to bring them to the same scale with non-intercalated phase.}
	\label{charge}
\end{figure*}

While our findings are largely consistent with key experimental data, the theory does not take into
  account the electron-phonon interaction, and we cannot exclude the possibility that it can also play some role in the
  enhancement of T$_{c}$ on intercalation.  As a hint towards addressing this question, we compute the charge
  susceptibilities in both QS\emph{GW} (within RPA approximation) and also with local vertex corrections from DMFT.  We
  find that intercalation causes the real part of the charge susceptibility $\chi^c$ to drop significantly within either
  approximation: Fig.~\ref{charge} (a-c) show the vertex corrected $\chi^c$ calculated from DMFT.  (Note that $\chi^c$ in the
  f-d phase is multiplied a factor of three to bring them to the same scale.)  A strong suppression of $\chi^c$ in the
  f-d phase suggests that collective charge excitations themselves can not drive pairing, nevertheless, if $\chi^c$
  becomes small enough ($\sim$0), then $\chi^{-1}$ can diverge. The electron phonon vertex is linear in $\chi^{-1}$ and
  can diverge too.  This suggests that while the electron-phonon plays little role in intrinsic FeSe, it may play some role
  in the intercalated case. A recent study explores the role of charge criticality in FeSe$_{1-x}$Te$_{x}$~\cite{mukasa2022enhanced}. It is likely that a charge mechanism, as elucidated above, is at play in doped and intercalated variants. A more definitive answer is beyond the scope of our present study.

To summarise, we perform \emph{ab-initio} all Green's function calculation for non-intercalated and intercalated FeSe in
the presence of vertex corrections.  QS\emph{GW} supplies a good reference one-body hamiltonian; this
  with the dynamical local self-energy and vertex from DMFT yields a very good \emph{ab-initio} description of the spin
  susceptibility~\cite{nematicity} and charge- and spin-fluctuation- mediated superconductivity. The vertex functions,
  with their orbital, momentum and energy dependence are directly computed out of the theory and no form factor is
  assumed.  We rigorously establish that the essential component of the superconducting pairing vertex is the magnetic
  vertex in FeSe.  Changes in the electronic density of states cannot explain the enhancements to T$_c$ in these
  systems. In the absence of the vertex, superconductivity in intercalated materials either would not increase or get
  weakly suppressed.

Intercalation enhances the superconducting instability  by a factor of five, primarily because the magnetic vertex gets enhanced by a similar factor at some
  particular $\bf q$ vectors in the Fe-d$_{xy}$ channel. The d$_{yz,xz}$ channels are also enhanced
  by a factor of two, but they always remain the secondary source of pairing glue. Such clear
orbital differential in two-particle channels is a hallmark of large Hund's coupling.

Further, we show that incorporating the effects of reduced electronic screening due to enhanced layer-separation
post-intercalation is crucial when constructing a realistic many-body Hamiltonian for these intercalated materials and
it is the enhancement in pairing vertex driven by such reduced electronic screening that can account for about a third
of the total enhancement in T$_{c}$. We believe, our work establishes the foundation for tetragonal FeSe where similarly
enhanced superconducting T$_{c}$ is realised on intercalation, alkali-doping, under-pressure, on ionic gating and
surface doping. Finally, we address the outstanding problem of missing d$_{xy}$ hole pocket from Fermi surfaces in bulk
FeSe. We show by creating an artificial potential shift to the d$_{xy}$ state, that even in the extreme case when the
d$_{xy}$ band energy is pushed nearly 100 meV below the Fermi energy, on intercalation, the pairing instability still
enhances by nearly a factor of 4. We also show that on intercalation,
  electronic spectral weight gets redistributed over various $\bf q$ vectors and in such cases, ${1}/{(T_{1}T)}$ may not
  enhance significantly above T$_{c}$ compared to the non-intercalated variant, in reasonable
    agreement with NMR measurements\cite{nmr}. Finally, we
    show that while \textbf{q}-selective enhancements in the pairing vertex are closely connected to enhancements in the
    magnetic susceptibility and concomitant superconducting instability, intercalation also induces a significant
    softening in collective charge excitations.  This raises the possibility that electron-phonon coupling may also
    contribute to superconductivity in intercalated FeSe.

\section*{Methods}
Single particle calculations (LDA, and energy band calculations with the static quasiparticlized QS\emph{GW} self-energy
$\Sigma^{0}(k)$) were performed on a 16$\times$16$\times$16 \emph{k}-mesh while the (relatively smooth) dynamical
self-energy $\Sigma(k)$ was constructed using a 8$\times$8$\times$8 \emph{k}-mesh and $\Sigma^{0}$(k)
is extracted from it.  The charge density was made self-consistent through iteration in the QS\emph{GW}
self-consistency cycle: it was iterated until the root mean square change in $\Sigma^{0}$ reached
10$^{-5}$\,Ry.  Thus the calculation was self-consistent in both $\Sigma^{0}(k)$ and the density.  At the end of
QS\emph{GW} cycles, we use the quasi-particlised electronic band structures as the starting point of our DMFT
calculations. The impurity Hamiltonian is solved with continuous time Quantum Monte Carlo
solver~\cite{haule,werner}. For projectors onto the Fe d subspace, we used projectors onto augmentation spheres,
following the method described in this reference~\cite{haule1}. This approach is sightly different from the approach used to compute U,J parameters in a previous work by Miyake et al.~\cite{miyake2010comparison}. Further, those numbers~\cite{miyake2010comparison} are computed while building the Hubbard Hamiltonian on top a DFT bath, while ours is a QS\emph{GW} bath. QS\emph{GW} already takes into long-range charge correlations missing from DFT, so it is only natural that the correlations (mostly of spin fluctuations origin) that our QS\emph{GW} calculations miss out would be lesser compared to DFT, leading to smaller U,J estimations. The double counting correlations are implemented using fully localised
limit approximation.  The DMFT for the dynamical self energy is iterated, and converges in 30 iterations. Calculations
for the single particle response functions are performed with 10$^{9}$ QMC steps per core and the statistics is averaged
over 128 cores. The two particle Green’s functions are sampled over a larger number of cores (40000-50000) to improve
the statistical error bars.  The local effective interactions for the correlated impurity Hamiltonian are given by
\emph{U} and \emph{J}. These are calculated within the constrained RPA~\cite{ferdi} from the QS\emph{GW} Hamiltonian
using an approach~\cite{swag19} similar to that of Ref~\cite{christoph}, using projectors from Ref.~\cite{haule1}.

\section*{Data Availability}

All input/output data can be made available on reasonable request. All the input file structures and the command lines to launch calculations are rigorously explained in the tutorials available on the Questaal webpage~\cite{questaal_web} \href{https://www.questaal.org/get/}.

\section*{Code Availability}
The source codes for LDA, QS\emph{GW} and QS$G\widehat{W}$ are available from~\cite{questaal_web}  \href{https://www.questaal.org/get/}  under the terms of the AGPLv3 license.

\section*{Acknowledgements}

MIK and SA are supported by the ERC Synergy Grant, project 854843 FASTCORR (Ultrafast dynamics of correlated electrons
in solids).  MvS (and SA in the late stages of this work) were supported by the U.S. Department of Energy, Office of
Science, Basic Energy Sciences, Division of Materials, under Contract No. DE-AC36-08GO28308.  SA acknowledges
discussions with Machteld E. Kamminga that inspired this work. We acknowledge PRACE for awarding us access to Irene-Rome
hosted by TGCC, France and Juwels Booster and Cluster, Germany. This work was also partly carried out on the Dutch
national e-infrastructure with the support of SURF Cooperative.  Late stages of calculations were performed using
computational resources sponsored by the Department of Energy: the Eagle facility at NREL, sponsored by the Office of
Energy Efficiency and also the National Energy Research Scientific Computing Center, under Contract
No. DE-AC02-05CH11231 using NERSC award BES-ERCAP0021783.

\section*{Author Contributions}
SA conceived the main theme of the work and performed the calculations. All authors have contributed to the writing of the paper and the analysis of the data.

\section*{Competing interests}
The authors declare no competing financial or non-financial interests.
\section*{Correspondence}
All correspondence, code and data requests should be made to SA.


\begin{thebibliography}{100}
	
	\bibitem{layeredcuo}
	R.~J. Cava, B.~Batlogg, J.~Krajewski, L.~Rupp, L.~Schneemeyer, T.~Siegrist,
	R.~VanDover, P.~Marsh, W.~Peck, P.~Gallagher, {\em et~al.},
	``Superconductivity near 70 k in a new family of layered copper oxides,''
	{\em Nature}, vol.~336, no.~6196, pp.~211--214, 1988.
	
	\bibitem{layered2d}
	W.~Zhang, Q.~Wang, Y.~Chen, Z.~Wang, and A.~T. Wee, ``Van der waals stacked 2d
	layered materials for optoelectronics,'' {\em 2D Materials}, vol.~3, no.~2,
	p.~022001, 2016.
	
	\bibitem{layeredgraphene}
	A.~K. Geim, ``Graphene: status and prospects,'' {\em Science}, vol.~324,
	no.~5934, pp.~1530--1534, 2009.
	
	\bibitem{layeredgr}
	K.~S. Novoselov, A.~K. Geim, S.~V. Morozov, D.~Jiang, M.~I. Katsnelson,
	I.~Grigorieva, S.~Dubonos, and a.~Firsov, ``Two-dimensional gas of massless
	dirac fermions in graphene,'' {\em Nature}, vol.~438, no.~7065, pp.~197--200,
	2005.
	
	\bibitem{layeredibs}
	Y.~Kamihara, T.~Watanabe, M.~Hirano, and H.~Hosono, ``Iron-based layered
	superconductor la [o1-x f x] feas (x= 0.05- 0.12) with t c= 26 k,'' {\em
		Journal of the American Chemical Society}, vol.~130, no.~11, pp.~3296--3297,
	2008.
	
	\bibitem{Kats_book}
	M.~I. Katsnelson, {\em The Physics of Graphene}.
	\newblock Cambridge University Press, 2020.
	
	\bibitem{2dmater_book}
	P.~Avouris, T.~F. Heinz, and T.~Low, eds., {\em 2D Materials: Properties and
		Devices}.
	\newblock Cambridge University Press, 2017.
	
	\bibitem{intercalated1}
	M.~S. Dresselhaus, ``Intercalation in layered materials,'' {\em MRS Bulletin},
	vol.~12, no.~3, pp.~24--28, 1987.
	
	\bibitem{intercalated2}
	F.~A. L{\'e}vy, {\em Intercalated layered materials}, vol.~6.
	\newblock Springer Science \& Business Media, 2012.
	
	\bibitem{intercalated3}
	M.~S. Stark, K.~L. Kuntz, S.~J. Martens, and S.~C. Warren, ``Intercalation of
	layered materials from bulk to 2d,'' {\em Advanced Materials}, vol.~31,
	no.~27, p.~1808213, 2019.
	
	\bibitem{exfo}
	K.~S. Novoselov, D.~Jiang, F.~Schedin, T.~Booth, V.~Khotkevich, S.~Morozov, and
	A.~K. Geim, ``Two-dimensional atomic crystals,'' {\em Proceedings of the
		National Academy of Sciences}, vol.~102, no.~30, pp.~10451--10453, 2005.
	
	\bibitem{exfoliation1}
	V.~Nicolosi, M.~Chhowalla, M.~G. Kanatzidis, M.~S. Strano, and J.~N. Coleman,
	``Liquid exfoliation of layered materials,'' {\em Science}, vol.~340,
	no.~6139, p.~1226419, 2013.
	
	\bibitem{exfoliation2}
	J.~N. Coleman, M.~Lotya, A.~O’Neill, S.~D. Bergin, P.~J. King, U.~Khan,
	K.~Young, A.~Gaucher, S.~De, R.~J. Smith, {\em et~al.}, ``Two-dimensional
	nanosheets produced by liquid exfoliation of layered materials,'' {\em
		Science}, vol.~331, no.~6017, pp.~568--571, 2011.
	
	\bibitem{exfoliation3}
	U.~Halim, C.~R. Zheng, Y.~Chen, Z.~Lin, S.~Jiang, R.~Cheng, Y.~Huang, and
	X.~Duan, ``A rational design of cosolvent exfoliation of layered materials by
	directly probing liquid--solid interaction,'' {\em Nature communications},
	vol.~4, no.~1, pp.~1--7, 2013.
	
	\bibitem{exfoliation4}
	A.~Ambrosi and M.~Pumera, ``Exfoliation of layered materials using
	electrochemistry,'' {\em Chemical Society Reviews}, vol.~47, no.~19,
	pp.~7213--7224, 2018.
	
	\bibitem{ibs}
	Y.~Kamihara, H.~Hiramatsu, M.~Hirano, R.~Kawamura, H.~Yanagi, T.~Kamiya, and
	H.~Hosono, ``Iron-based layered superconductor: Laofep,'' {\em Journal of the
		American Chemical Society}, vol.~128, no.~31, pp.~10012--10013, 2006.
	
	\bibitem{mcqueen}
	T.~M. McQueen, A.~J. Williams, P.~W. Stephens, J.~Tao, Y.~Zhu, V.~Ksenofontov,
	F.~Casper, C.~Felser, and R.~J. Cava, ``Tetragonal-to-orthorhombic structural
	phase transition at 90 k in the superconductor
	${\mathrm{fe}}_{1.01}\mathrm{Se}$,'' {\em Phys. Rev. Lett.}, vol.~103,
	p.~057002, Jul 2009.
	
	\bibitem{mizuguchi}
	Y.~Mizuguchi, F.~Tomioka, S.~Tsuda, T.~Yamaguchi, and Y.~Takano,
	``Superconductivity at 27 k in tetragonal fese under high pressure,'' {\em
		Applied Physics Letters}, vol.~93, no.~15, p.~152505, 2008.
	
	\bibitem{shipra}
	R.~Shipra, H.~Takeya, K.~Hirata, and A.~Sundaresan, ``Effects of ni and co
	doping on the physical properties of tetragonal fese0. 5te0. 5
	superconductor,'' {\em Physica C: Superconductivity}, vol.~470, no.~13-14,
	pp.~528--532, 2010.
	
	\bibitem{galluzzi}
	A.~Galluzzi, M.~Polichetti, K.~Buchkov, E.~Nazarova, D.~Mancusi, and S.~Pace,
	``Critical current and flux dynamics in ag-doped fese superconductor,'' {\em
		Superconductor Science and Technology}, vol.~30, no.~2, p.~025013, 2016.
	
	\bibitem{sun2017}
	F.~Sun, Z.~Guo, H.~Zhang, and W.~Yuan, ``S/te co-doping in tetragonal fese with
	unchanged lattice parameters: Effects on superconductivity and electronic
	structure,'' {\em Journal of Alloys and Compounds}, vol.~700, pp.~43--48,
	2017.
	
	\bibitem{craco2014}
	L.~Craco, M.~Laad, and S.~Leoni, ``Normal-state correlated electronic structure
	of tetragonal fese superconductor,'' in {\em Journal of Physics: Conference
		Series}, vol.~487, p.~012017, IOP Publishing, 2014.
	
	\bibitem{pressure}
	T.~Imai, K.~Ahilan, F.~L. Ning, T.~M. McQueen, and R.~J. Cava, ``Why does
	undoped fese become a high-${T}_{c}$ superconductor under pressure?,'' {\em
		Phys. Rev. Lett.}, vol.~102, p.~177005, Apr 2009.
	
	\bibitem{pressure1}
	S.~Medvedev, T.~McQueen, I.~Troyan, T.~Palasyuk, M.~Eremets, R.~Cava,
	S.~Naghavi, F.~Casper, V.~Ksenofontov, G.~Wortmann, {\em et~al.},
	``Electronic and magnetic phase diagram of $\beta$-fe1. 01se with
	superconductivity at 36.7 k under pressure,'' {\em Nature materials}, vol.~8,
	no.~8, pp.~630--633, 2009.
	
	\bibitem{pressure2}
	Y.~Mizuguchi, F.~Tomioka, S.~Tsuda, T.~Yamaguchi, and Y.~Takano,
	``Superconductivity at 27 k in tetragonal fese under high pressure,'' {\em
		Applied Physics Letters}, vol.~93, no.~15, p.~152505, 2008.
	
	\bibitem{qing}
	W.~Qing-Yan, L.~Zhi, Z.~Wen-Hao, Z.~Zuo-Cheng, Z.~Jin-Song, L.~Wei, D.~Hao,
	O.~Yun-Bo, D.~Peng, C.~Kai, {\em et~al.}, ``Interface-induced
	high-temperature superconductivity in single unit-cell fese films on
	srtio3,'' {\em Chinese Physics Letters}, vol.~29, no.~3, p.~037402, 2012.
	
	\bibitem{ge2014}
	J.-F. Ge, Z.-L. Liu, C.~Liu, C.-L. Gao, D.~Qian, Q.-K. Xue, Y.~Liu, and J.-F.
	Jia, ``Superconductivity above 100 k in single-layer fese films on doped
	srtio 3,'' {\em Nature materials}, vol.~14, no.~3, p.~285, 2015.
	
	\bibitem{noji}
	T.~Noji, T.~Hatakeda, S.~Hosono, T.~Kawamata, M.~Kato, and Y.~Koike,
	``Synthesis and post-annealing effects of
	alkaline-metal-ethylenediamine-intercalated superconductors ax (c2h8n2) yfe2-
	zse2 (a= li, na) with tc= 45 k,'' {\em Physica C: Superconductivity and its
		Applications}, vol.~504, pp.~8--11, 2014.
	
	\bibitem{wang}
	Z.~Wang, Y.~Song, H.~Shi, Z.~Wang, Z.~Chen, H.~Tian, G.~Chen, J.~Guo, H.~Yang,
	and J.~Li, ``Microstructure and ordering of iron vacancies in the
	superconductor system k y fe x se 2 as seen via transmission electron
	microscopy,'' {\em Physical Review B}, vol.~83, no.~14, p.~140505, 2011.
	
	\bibitem{potassium}
	A.-m. Zhang, T.-l. Xia, K.~Liu, W.~Tong, Z.-r. Yang, and Q.-m. Zhang,
	``Superconductivity at 44 k in k intercalated fese system with excess fe,''
	{\em Scientific reports}, vol.~3, no.~1, pp.~1--5, 2013.
	
	\bibitem{barium}
	K.~Yusenko, J.~Sottmann, H.~Emerich, W.~Crichton, L.~Malavasi, and
	S.~Margadonna, ``Hyper-expanded interlayer separations in superconducting
	barium intercalates of fese,'' {\em Chemical Communications}, vol.~51,
	no.~33, pp.~7112--7115, 2015.
	
	\bibitem{sodium}
	T.~Ying, X.~Chen, G.~Wang, S.~Jin, T.~Zhou, X.~Lai, H.~Zhang, and W.~Wang,
	``Observation of superconductivity at 30~ 46k in axfe2se2 (a= li, na, ba, sr,
	ca, yb and eu),'' {\em Scientific Reports}, vol.~2, no.~1, pp.~1--7, 2012.
	
	\bibitem{surface1}
	Y.~Miyata, K.~Nakayama, K.~Sugawara, T.~Sato, and T.~Takahashi,
	``High-temperature superconductivity in potassium-coated multilayer fese thin
	films,'' {\em Nature materials}, vol.~14, no.~8, pp.~775--779, 2015.
	
	\bibitem{ionic1}
	K.~Hanzawa, H.~Sato, H.~Hiramatsu, T.~Kamiya, and H.~Hosono, ``Electric
	field-induced superconducting transition of insulating fese thin film at 35
	k,'' {\em Proceedings of the National Academy of Sciences}, vol.~113, no.~15,
	pp.~3986--3990, 2016.
	
	\bibitem{ionic2}
	B.~Lei, J.~Cui, Z.~Xiang, C.~Shang, N.~Wang, G.~Ye, X.~Luo, T.~Wu, Z.~Sun, and
	X.~Chen, ``Evolution of high-temperature superconductivity from a low-t c
	phase tuned by carrier concentration in fese thin flakes,'' {\em Physical
		review letters}, vol.~116, no.~7, p.~077002, 2016.
	
	\bibitem{ionic3}
	X.~Lu, N.~Wang, H.~Wu, Y.~Wu, D.~Zhao, X.~Zeng, X.~Luo, T.~Wu, W.~Bao,
	G.~Zhang, {\em et~al.}, ``Coexistence of superconductivity and
	antiferromagnetism in (li0. 8fe0. 2) ohfese,'' {\em Nature materials},
	vol.~14, no.~3, pp.~325--329, 2015.
	
	\bibitem{ionic4}
	J.~Shiogai, Y.~Ito, T.~Mitsuhashi, T.~Nojima, and A.~Tsukazaki,
	``Electric-field-induced superconductivity in electrochemically etched
	ultrathin fese films on srtio3 and mgo,'' {\em Nature Physics}, vol.~12,
	no.~1, pp.~42--46, 2016.
	
	\bibitem{sedlmaier1}
	S.~J. Sedlmaier, S.~J. Cassidy, R.~G. Morris, M.~Drakopoulos, C.~Reinhard,
	S.~J. Moorhouse, D.~O’Hare, P.~Manuel, D.~Khalyavin, and S.~J. Clarke,
	``Ammonia-rich high-temperature superconducting intercalates of iron selenide
	revealed through time-resolved in situ x-ray and neutron diffraction,'' {\em
		Journal of the American Chemical Society}, vol.~136, no.~2, pp.~630--633,
	2014.
	
	\bibitem{sedlmaier2}
	M.~Burrard-Lucas, D.~G. Free, S.~J. Sedlmaier, J.~D. Wright, S.~J. Cassidy,
	Y.~Hara, A.~J. Corkett, T.~Lancaster, P.~J. Baker, S.~J. Blundell, {\em
		et~al.}, ``Enhancement of the superconducting transition temperature of fese
	by intercalation of a molecular spacer layer,'' {\em Nature materials},
	vol.~12, no.~1, pp.~15--19, 2013.
	
	\bibitem{kamminga}
	M.~E. Kamminga, S.~J. Cassidy, P.~P. Jana, M.~Elgaml, N.~D. Kelly, and S.~J.
	Clarke, ``Intercalates of bi 2 se 3 studied in situ by time-resolved powder
	x-ray diffraction and neutron diffraction,'' {\em Dalton Transactions},
	vol.~50, no.~33, pp.~11376--11379, 2021.
	
	\bibitem{coldea}
	A.~I. Coldea and M.~D. Watson, ``The key ingredients of the electronic
	structure of fese,'' {\em Annual Review of Condensed Matter Physics}, vol.~9,
	pp.~125--146, 2018.
	
	\bibitem{yin}
	Z.~P. Yin, K.~Haule, and G.~Kotliar, ``Spin dynamics and orbital-antiphase
	pairing symmetry in iron-based superconductors,'' {\em Nature Physics},
	vol.~10, no.~11, pp.~845--850, 2014.
	
	\bibitem{symmetry2021}
	S.~Acharya, D.~Pashov, F.~Jamet, and M.~van Schilfgaarde, ``Electronic origin
	of tc in bulk and monolayer fese,'' {\em Symmetry}, vol.~13, no.~2, p.~169,
	2021.
	
	\bibitem{gretarsson}
	H.~Gretarsson, A.~Lupascu, J.~Kim, D.~Casa, T.~Gog, W.~Wu, S.~Julian, Z.~Xu,
	J.~Wen, G.~Gu, {\em et~al.}, ``Revealing the dual nature of magnetism in iron
	pnictides and iron chalcogenides using x-ray emission spectroscopy,'' {\em
		Physical Review B}, vol.~84, no.~10, p.~100509, 2011.
	
	\bibitem{medici2013}
	A.~Georges, L.~d. Medici, and J.~Mravlje, ``Strong correlations from hund’s
	coupling,'' {\em Annual Review of Condensed Matter Physics}, vol.~4, no.~1,
	pp.~137--178, 2013.
	
	\bibitem{haule2009coherence}
	K.~Haule and G.~Kotliar, ``Coherence--incoherence crossover in the normal state
	of iron oxypnictides and importance of hund's rule coupling,'' {\em New
		journal of physics}, vol.~11, no.~2, p.~025021, 2009.
	
	\bibitem{nematicity}
	S.~Acharya, D.~Pashov, and M.~van Schilfgaarde, ``Role of nematicity in
	controlling spin fluctuations and superconducting ${T}_{c}$ in bulk fese,''
	{\em Phys. Rev. B}, vol.~105, p.~144507, Apr 2022.
	
	\bibitem{valenti}
	D.~Guterding, H.~O. Jeschke, P.~Hirschfeld, and R.~Valent{\'\i}, ``Unified
	picture of the doping dependence of superconducting transition temperatures
	in alkali metal/ammonia intercalated fese,'' {\em Physical Review B},
	vol.~91, no.~4, p.~041112, 2015.
	
	\bibitem{bcs}
	J.~Bardeen, L.~N. Cooper, and J.~R. Schrieffer, ``Theory of
	superconductivity,'' {\em Phys. Rev.}, vol.~108, pp.~1175--1204, Dec 1957.
	
	\bibitem{VIK_book}
	S.~V. Vonsovsky, Y.~A. Izyumov, and E.~Z. Kurmaev, {\em Superconductivity of
		Transition Metals, Their Alloys and Compounds}.
	\newblock Springer-Verlag, 1982.
	
	\bibitem{HgBaCuO}
	D.~L. Novikov, M.~I. Katsnelson, J.~Yu, A.~V. Postnikov, and A.~J. Freeman,
	``Pressure-induced phonon softening and electronic topological transition in
	hgba$_2$cuo$_4$,'' {\em Phys. Rev. B}, vol.~54, pp.~1313--1319, July 1996.
	
	\bibitem{vanhove}
	R.~S. Markiewicz, ``A survey of the van hove scenario for high-tc
	superconductivity with special emphasis on pseudogaps and striped phases,''
	{\em J. Phys. Chem. Solids}, vol.~58, pp.~1179--1310, August 1997.
	
	\bibitem{kotani}
	T.~Kotani, M.~van Schilfgaarde, and S.~V. Faleev, ``Quasiparticle
	self-consistent $gw$ method: A basis for the independent-particle
	approximation,'' {\em Phys. Rev. B}, vol.~76, p.~165106, Oct 2007.
	
	\bibitem{questaal-paper}
	D.~Pashov, S.~Acharya, W.~R. Lambrecht, J.~Jackson, K.~D. Belashchenko,
	A.~Chantis, F.~Jamet, and M.~van Schilfgaarde, ``Questaal: a package of
	electronic structure methods based on the linear muffin-tin orbital
	technique,'' {\em Computer Physics Communications}, vol.~249, p.~107065,
	2020.
	
	\bibitem{dmft1}
	S.~L. Skornyakov, V.~I. Anisimov, D.~Vollhardt, and I.~Leonov, ``Effect of
	electron correlations on the electronic structure and phase stability of fese
	upon lattice expansion,'' {\em Phys. Rev. B}, vol.~96, p.~035137, Jul 2017.
	
	\bibitem{dmft2}
	S.~Mandal, R.~E. Cohen, and K.~Haule, ``Strong pressure-dependent
	electron-phonon coupling in fese,'' {\em Physical Review B}, vol.~89, no.~22,
	p.~220502, 2014.
	
	\bibitem{dmft3}
	S.~Mandal, P.~Zhang, S.~Ismail-Beigi, and K.~Haule, ``How correlated is the
	fese/srtio 3 system?,'' {\em Physical review letters}, vol.~119, no.~6,
	p.~067004, 2017.
	
	\bibitem{dmft4}
	M.~Aichhorn, S.~Biermann, T.~Miyake, A.~Georges, and M.~Imada, ``Theoretical
	evidence for strong correlations and incoherent metallic state in fese,''
	{\em Physical Review B}, vol.~82, no.~6, p.~064504, 2010.
	
	\bibitem{dmft5}
	M.~D. Watson, S.~Backes, A.~A. Haghighirad, M.~Hoesch, T.~K. Kim, A.~I. Coldea,
	and R.~Valent\'{\i}, ``Formation of hubbard-like bands as a fingerprint of
	strong electron-electron interactions in fese,'' {\em Phys. Rev. B}, vol.~95,
	p.~081106, Feb 2017.
	
	\bibitem{ref16}
	C.-H. Lee, A.~Iyo, H.~Eisaki, H.~Kito, M.~Teresa Fernandez-Diaz, T.~Ito,
	K.~Kihou, H.~Matsuhata, M.~Braden, and K.~Yamada, ``Effect of structural
	parameters on superconductivity in fluorine-free lnfeaso1-y (ln= la, nd),''
	{\em Journal of the Physical Society of Japan}, vol.~77, no.~8, p.~083704,
	2008.
	
	\bibitem{ref17}
	H.~Hosono and K.~Kuroki, ``Iron-based superconductors: Current status of
	materials and pairing mechanism,'' {\em Physica C: Superconductivity and its
		Applications}, vol.~514, pp.~399--422, 2015.
	
	\bibitem{ref18}
	M.~Yi, Y.~Zhang, Z.-X. Shen, and D.~Lu, ``Role of the orbital degree of freedom
	in iron-based superconductors,'' {\em npj Quantum Materials}, vol.~2, no.~1,
	pp.~1--12, 2017.
	
	\bibitem{ref19}
	T.~Shibauchi, T.~Hanaguri, and Y.~Matsuda, ``Exotic superconducting states in
	fese-based materials,'' {\em Journal of the Physical Society of Japan},
	vol.~89, no.~10, p.~102002, 2020.
	
	\bibitem{ref20}
	M.~Kitatani, T.~Sch{\"a}fer, H.~Aoki, and K.~Held, ``Why the critical
	temperature of high-t c cuprate superconductors is so low: The importance of
	the dynamical vertex structure,'' {\em Physical Review B}, vol.~99, no.~4,
	p.~041115, 2019.
	
	\bibitem{ref21}
	K.~Held, L.~Si, P.~Worm, O.~Janson, R.~Arita, Z.~Zhong, J.~M. Tomczak, and
	M.~Kitatani, ``Phase diagram of nickelate superconductors calculated by
	dynamical vertex approximation,'' {\em Frontiers in Physics}, p.~803, 2022.
	
	\bibitem{swag19}
	S.~Acharya, D.~Pashov, C.~Weber, H.~Park, L.~Sponza, and M.~Van~Schilfgaarde,
	``Evening out the spin and charge parity to increase tc in sr2ruo4,'' {\em
		Communications Physics}, vol.~2, no.~1, pp.~1--8, 2019.
	
	\bibitem{prl2020}
	S.~Acharya, D.~Pashov, F.~Jamet, and M.~van Schilfgaarde, ``Controlling
	${\mathit{t}}_{c}$ through band structure and correlation engineering in
	collapsed and uncollapsed phases of iron arsenides,'' {\em Phys. Rev. Lett.},
	vol.~124, p.~237001, Jun 2020.
	
	\bibitem{kostin2018}
	A.~Kostin, P.~O. Sprau, A.~Kreisel, Y.~X. Chong, A.~E. B{\"o}hmer, P.~C.
	Canfield, P.~J. Hirschfeld, B.~M. Andersen, and J.~S. Davis, ``Imaging
	orbital-selective quasiparticles in the hund’s metal state of fese,'' {\em
		Nature materials}, p.~1, 2018.
	
	\bibitem{watsonprb}
	M.~D. Watson, T.~K. Kim, A.~A. Haghighirad, N.~R. Davies, A.~McCollam,
	A.~Narayanan, S.~F. Blake, Y.~L. Chen, S.~Ghannadzadeh, A.~J. Schofield,
	M.~Hoesch, C.~Meingast, T.~Wolf, and A.~I. Coldea, ``Emergence of the nematic
	electronic state in fese,'' {\em Phys. Rev. B}, vol.~91, p.~155106, Apr 2015.
	
	\bibitem{watsonnjp}
	M.~D. Watson, A.~A. Haghighirad, L.~C. Rhodes, M.~Hoesch, and T.~K. Kim,
	``Electronic anisotropies revealed by detwinned angle-resolved photo-emission
	spectroscopy measurements of fese,'' {\em New Journal of Physics}, vol.~19,
	no.~10, p.~103021, 2017.
	
	\bibitem{ref11}
	T.~Qian, X.-P. Wang, W.-C. Jin, P.~Zhang, P.~Richard, G.~Xu, X.~Dai, Z.~Fang,
	J.-G. Guo, X.-L. Chen, {\em et~al.}, ``Absence of a holelike fermi surface
	for the iron-based k 0.8 fe 1.7 se 2 superconductor revealed by
	angle-resolved photoemission spectroscopy,'' {\em Physical review letters},
	vol.~106, no.~18, p.~187001, 2011.
	
	\bibitem{ref12}
	S.~Rinott, K.~Chashka, A.~Ribak, E.~D. Rienks, A.~Taleb-Ibrahimi, P.~Le~Fevre,
	F.~Bertran, M.~Randeria, and A.~Kanigel, ``Tuning across the bcs-bec
	crossover in the multiband superconductor fe1+ y se x te1- x: An
	angle-resolved photoemission study,'' {\em Science advances}, vol.~3, no.~4,
	p.~e1602372, 2017.
	
	\bibitem{ref13}
	K.~Matsumoto, D.~Ogura, and K.~Kuroki, ``Wide applicability of high-t c pairing
	originating from coexisting wide and incipient narrow bands in
	quasi-one-dimensional systems,'' {\em Physical Review B}, vol.~97, no.~1,
	p.~014516, 2018.
	
	\bibitem{ref14}
	K.~Matsumoto, D.~Ogura, and K.~Kuroki, ``Strongly enhanced superconductivity
	due to finite energy spin fluctuations induced by an incipient band: A flex
	study on the bilayer hubbard model with vertical and diagonal interlayer
	hoppings,'' {\em Journal of the Physical Society of Japan}, vol.~89, no.~4,
	p.~044709, 2020.
	
	\bibitem{ref15}
	K.~Kuroki, T.~Higashida, and R.~Arita, ``High-${T}_{c}$ superconductivity due
	to coexisting wide and narrow bands: A fluctuation exchange study of the
	hubbard ladder as a test case,'' {\em Phys. Rev. B}, vol.~72, p.~212509, Dec
	2005.
	
	\bibitem{mik1}
	M.~Katsnelson and A.~Trefilov, ``Anomalies in properties of metals and alloys
	due to electron correlations,'' {\em Physics Letters A}, vol.~109, no.~3,
	pp.~109--112, 1985.
	
	\bibitem{mik2}
	M.~Katsnelson and A.~Trefilov, ``Anomalies of electronic and lattice properties
	of metals and alloys, caused by screening anomalies,'' {\em Physica B:
		Condensed Matter}, vol.~163, no.~1-3, pp.~182--184, 1990.
	
	\bibitem{arpes}
	L.~Zhao, A.~Liang, D.~Yuan, Y.~Hu, D.~Liu, J.~Huang, S.~He, B.~Shen, Y.~Xu,
	X.~Liu, {\em et~al.}, ``Common electronic origin of superconductivity in (li,
	fe) ohfese bulk superconductor and single-layer fese/srtio3 films,'' {\em
		Nature communications}, vol.~7, no.~1, p.~10608, 2016.
	
	\bibitem{watson2016}
	M.~Watson, T.~Kim, L.~Rhodes, M.~Eschrig, M.~Hoesch, A.~Haghighirad, and
	A.~Coldea, ``Evidence for unidirectional nematic bond ordering in fese,''
	{\em Physical Review B}, vol.~94, no.~20, p.~201107, 2016.
	
	\bibitem{watson2017}
	M.~D. Watson, A.~A. Haghighirad, L.~C. Rhodes, M.~Hoesch, and T.~K. Kim,
	``Electronic anisotropies revealed by detwinned angle-resolved photo-emission
	spectroscopy measurements of fese,'' {\em New Journal of Physics}, vol.~19,
	no.~10, p.~103021, 2017.
	
	\bibitem{imai2009}
	T.~Imai, K.~Ahilan, F.~Ning, T.~McQueen, and R.~J. Cava, ``Why does undoped
	fese become a high-t c superconductor under pressure?,'' {\em Physical review
		letters}, vol.~102, no.~17, p.~177005, 2009.
	
	\bibitem{alkaliarpes1}
	Y.~Zhang, L.~Yang, M.~Xu, Z.~Ye, F.~Chen, C.~He, H.~Xu, J.~Jiang, B.~Xie,
	J.~Ying, {\em et~al.}, ``Nodeless superconducting gap in a x fe2se2 (a= k,
	cs) revealed by angle-resolved photoemission spectroscopy,'' {\em Nature
		materials}, vol.~10, no.~4, pp.~273--277, 2011.
	
	\bibitem{alkaliarpes2}
	T.~Qian, X.-P. Wang, W.-C. Jin, P.~Zhang, P.~Richard, G.~Xu, X.~Dai, Z.~Fang,
	J.-G. Guo, X.-L. Chen, and H.~Ding, ``Absence of a holelike fermi surface for
	the iron-based ${\mathrm{k}}_{0.8}{\mathrm{fe}}_{1.7}{\mathrm{se}}_{2}$
	superconductor revealed by angle-resolved photoemission spectroscopy,'' {\em
		Phys. Rev. Lett.}, vol.~106, p.~187001, May 2011.
	
	\bibitem{alkaliarpes3}
	D.~Mou, S.~Liu, X.~Jia, J.~He, Y.~Peng, L.~Zhao, L.~Yu, G.~Liu, S.~He, X.~Dong,
	J.~Zhang, H.~Wang, C.~Dong, M.~Fang, X.~Wang, Q.~Peng, Z.~Wang, S.~Zhang,
	F.~Yang, Z.~Xu, C.~Chen, and X.~J. Zhou, ``Distinct fermi surface topology
	and nodeless superconducting gap in a
	$({\mathrm{tl}}_{0.58}{\mathrm{rb}}_{0.42}){\mathrm{fe}}_{1.72}{\mathrm{se}}_{2}$
	superconductor,'' {\em Phys. Rev. Lett.}, vol.~106, p.~107001, Mar 2011.
	
	\bibitem{incipient}
	A.~Linscheid, S.~Maiti, Y.~Wang, S.~Johnston, and P.~J. Hirschfeld, ``High
	${T}_{c}$ via spin fluctuations from incipient bands: Application to
	monolayers and intercalates of fese,'' {\em Phys. Rev. Lett.}, vol.~117,
	p.~077003, Aug 2016.
	
	\bibitem{nmr}
	M.~c. v.~M. Hrovat, P.~Jegli\ifmmode~\check{c}\else \v{c}\fi{},
	M.~Klanj\ifmmode~\check{s}\else \v{s}\fi{}ek, T.~Hatakeda, T.~Noji,
	Y.~Tanabe, T.~Urata, K.~K. Huynh, Y.~Koike, K.~Tanigaki, and
	D.~Ar\ifmmode~\check{c}\else \v{c}\fi{}on, ``Enhanced superconducting
	transition temperature in hyper-interlayer-expanded fese despite the
	suppressed electronic nematic order and spin fluctuations,'' {\em Phys. Rev.
		B}, vol.~92, p.~094513, Sep 2015.
	
	\bibitem{fesete}
	Z.~Xu, J.~Wen, G.~Xu, Q.~Jie, Z.~Lin, Q.~Li, S.~Chi, D.~K. Singh, G.~Gu, and
	J.~M. Tranquada, ``Disappearance of static magnetic order and evolution of
	spin fluctuations in
	${\text{fe}}_{1+\ensuremath{\delta}}{\text{se}}_{x}{\text{te}}_{1\ensuremath{-}x}$,''
	{\em Phys. Rev. B}, vol.~82, p.~104525, Sep 2010.
	
	\bibitem{alkali}
	J.~Guo, S.~Jin, G.~Wang, S.~Wang, K.~Zhu, T.~Zhou, M.~He, and X.~Chen,
	``Superconductivity in the iron selenide
	${\text{k}}_{x}{\text{fe}}_{2}{\text{se}}_{2}$
	$(0\ensuremath{\le}x\ensuremath{\le}1.0)$,'' {\em Phys. Rev. B}, vol.~82,
	p.~180520, Nov 2010.
	
	\bibitem{alkaliq}
	J.~T. Park, G.~Friemel, Y.~Li, J.-H. Kim, V.~Tsurkan, J.~Deisenhofer, H.-A.
	Krug~von Nidda, A.~Loidl, A.~Ivanov, B.~Keimer, and D.~S. Inosov, ``Magnetic
	resonant mode in the low-energy spin-excitation spectrum of superconducting
	${\mathrm{rb}}_{2}{\mathrm{fe}}_{4}{\mathrm{se}}_{5}$ single crystals,'' {\em
		Phys. Rev. Lett.}, vol.~107, p.~177005, Oct 2011.
	
	\bibitem{alkaliq1}
	G.~Friemel, J.~T. Park, T.~A. Maier, V.~Tsurkan, Y.~Li, J.~Deisenhofer, H.-A.
	Krug~von Nidda, A.~Loidl, A.~Ivanov, B.~Keimer, and D.~S. Inosov,
	``Reciprocal-space structure and dispersion of the magnetic resonant mode in
	the superconducting phase of rb${}_{x}$fe${}_{2\ensuremath{-}y}$se${}_{2}$
	single crystals,'' {\em Phys. Rev. B}, vol.~85, p.~140511, Apr 2012.
	
	\bibitem{alkalinmr}
	W.~Yu, L.~Ma, J.~B. He, D.~M. Wang, T.-L. Xia, G.~F. Chen, and W.~Bao,
	``$^{77}\mathrm{Se}$ nmr study of the pairing symmetry and the spin dynamics
	in ${\mathrm{k}}_{y}{\mathrm{fe}}_{2\ensuremath{-}x}{\mathrm{se}}_{2}$,''
	{\em Phys. Rev. Lett.}, vol.~106, p.~197001, May 2011.
	
	\bibitem{sr2ruo4}
	S.~Acharya, D.~Pashov, E.~Chachkarova, M.~v. Schilfgaarde, and C.~Weber,
	``Electronic structure correspondence of singlet-triplet scale separation in
	strained sr2ruo4,'' {\em Applied Sciences}, vol.~11, no.~2, p.~508, 2021.
	
	\bibitem{mukasa2022enhanced}
	K.~Mukasa, K.~Ishida, S.~Imajo, M.~Qiu, M.~Saito, K.~Matsuura, Y.~Sugimura,
	S.~Liu, Y.~Uezono, T.~Otsuka, {\em et~al.}, ``Enhanced superconducting
	pairing strength near a nonmagnetic nematic quantum critical point,'' {\em
		arXiv preprint arXiv:2202.11657}, 2022.
	
	\bibitem{haule}
	K.~Haule, ``Quantum monte carlo impurity solver for cluster dynamical
	mean-field theory and electronic structure calculations with adjustable
	cluster base,'' {\em Phys. Rev. B}, vol.~75, p.~155113, Apr 2007.
	
	\bibitem{werner}
	P.~Werner, A.~Comanac, L.~de' Medici, M.~Troyer, and A.~J. Millis,
	``Continuous-time solver for quantum impurity models,'' {\em Phys. Rev.
		Lett.}, vol.~97, p.~076405, Aug 2006.
	
	\bibitem{haule1}
	K.~Haule, C.-H. Yee, and K.~Kim, ``Dynamical mean-field theory within the
	full-potential methods: Electronic structure of ${\text{ceirin}}_{5}$,
	${\text{cecoin}}_{5}$, and ${\text{cerhin}}_{5}$,'' {\em Phys. Rev. B},
	vol.~81, p.~195107, May 2010.
	
	\bibitem{miyake2010comparison}
	T.~Miyake, K.~Nakamura, R.~Arita, and M.~Imada, ``Comparison of ab initio
	low-energy models for lafepo, lafeaso, bafe2as2, lifeas, fese, and fete:
	electron correlation and covalency,'' {\em Journal of the Physical Society of
		Japan}, vol.~79, no.~4, p.~044705, 2010.
	
	\bibitem{ferdi}
	T.~Miyake, F.~Aryasetiawan, and M.~Imada, ``Ab initio procedure for
	constructing effective models of correlated materials with entangled band
	structure,'' {\em Phys. Rev. B}, vol.~80, p.~155134, Oct 2009.
	
	\bibitem{christoph}
	E.~\ifmmode \mbox{\c{S}}\else \c{S}\fi{}a\ifmmode \mbox{\c{s}}\else
	\c{s}\fi{}\ifmmode \imath \else \i \fi{}o\ifmmode~\breve{g}\else
	\u{g}\fi{}lu, C.~Friedrich, and S.~Bl\"ugel, ``Effective coulomb interaction
	in transition metals from constrained random-phase approximation,'' {\em
		Phys. Rev. B}, vol.~83, p.~121101, Mar 2011.
	
	\bibitem{questaal_web}
	``Questaal website.'' \url{https://www.questaal.org}.
	
\end{thebibliography}

\section*{Figures}

\end{document}